\documentclass[final,3p,times]{elsarticle}

\usepackage{amssymb}
\usepackage{amsmath}
\usepackage{amsthm}
\usepackage{bm}
\usepackage{hyperref}
\usepackage{multirow}
\usepackage{graphicx,tabularx,multirow, color}
\usepackage{subfigure}

\theoremstyle{definition} 

\newcommand{\ud}{\mathrm{d}}
\newcommand{\un}{\mathrm{n}}
\newcommand{\ue}{\mathrm{e}}
\newcommand{\abs}[1]{\left|#1\right|}
\newcommand{\norm}[1]{\left|#1\right|}
\newcommand{\X}{\mathbf{x}}
\newcommand{\Y}{\mathbf{y}}
\newcommand{\J}{\mathbf{J}}
\newcommand{\U}{\mathbf{u}}
\newcommand{\fpt}{\X}%field point
\newcommand{\spt}{\Y}%source point
\newcommand{\pfrac}[2]{\frac{\partial #1}{\partial #2}}
\newcommand{\ppfrac}[2]{\frac{\partial^2 #1}{\partial #2^2}}
\newcommand{\ppfracd}[3]{\frac{\partial^2 #1}{\partial #2 \partial #3}}
\newcommand{\optrS}[1]{\int_{S}G(\fpt,\spt)#1(\spt)\ud S(\spt)}
\newcommand{\optrD}[1]{\int_{S}\pfrac{G(\fpt,\spt)}{\un(\spt)}#1(\spt)\ud S(\spt)}
\newcommand{\optrM}[1]{\int_{S}\pfrac{G(\fpt,\spt)}{\un(\fpt)}#1(\spt)\ud S(\spt)}
\newcommand{\optrH}[1]{\int_{S}\ppfracd{G(\fpt,\spt)}{\un(\fpt)}{\un(\spt)}#1(\spt)\ud S(\spt)}

\DeclareMathOperator{\opS}{\mathcal{S}}
\DeclareMathOperator{\opD}{\mathcal{D}}
\DeclareMathOperator{\opM}{\mathcal{M}}
\DeclareMathOperator{\opH}{\mathcal{H}}
\begin{document}

\begin{frontmatter}


\title{An efficient method for evaluating BEM singular integrals on curved elements with application in acoustic analysis}


\author[]{Junjie Rong}
\ead{rxrjj@126.com}
\author[]{Lihua Wen\corref{cor1}}
\ead{lhwen@nwpu.edu.cn}
\author[]{Jinyou Xiao}
\ead{xiaojy@nwpu.edu.cn}

\cortext[cor1]{Corresponding author}

\address{College of Astronautics, Northwestern Polytechnical University, Xi'an 710072, P. R. China}

\begin{abstract}

The polar coordinate transformation (PCT) method has been extensively used to treat various singular integrals in the boundary element method (BEM).  However, the resultant integrands of the PCT tend to become nearly singular when (1) the aspect ratio of the element is large or (2) the field point is closed to the element boundary; thus a large number of quadrature points are needed to achieve a relatively high accuracy. In this paper, the first problem is circumvented by using a conformal transformation so that the geometry of the curved physical element is preserved in the transformed domain. The second problem is alleviated by using a sigmoidal transformation, which makes the quadrature points more concentrated around the near singularity.

By combining the proposed two transformations with the Guiggiani's method in [M. Guiggiani, \emph{et al}.
 A general algorithm for the numerical solution of hypersingular boundary integral equations.
 \emph{ASME Journal of Applied Mechanics}, 59(1992), 604-614], one obtains an efficient and robust numerical method for computing the weakly-, strongly- and hyper-singular integrals in high-order BEM with curved elements. Numerical integration results show that, compared with the original PCT, the present method can reduce the number of quadrature points considerably, for given accuracy. For further verification, the method is incorporated into a $2$-order Nystr\"om BEM code for solving acoustic Burton-Miller boundary integral equation. It is shown that the method can retain the convergence rate of the BEM with much less quadrature points than the existing PCT. The method is implemented
in C language and freely available.

\end{abstract}

\begin{keyword}
singular integrals \sep boundary element method \sep Nystr\"om method \sep acoustics

\end{keyword}

\end{frontmatter}

\section{Introduction}
%The boundary element method (BEM) plays a dominate role in solving infinite and simi-infinite acoustic wave problems due to easy handling of boundary condition at infinity and dimension reduction However, the BEM based on the conventional boundary integral equation (CBIE) fails to yield unique solutions for exterior acoustic problems at the eigenfrequencies of the associated interior problems. A sound and effective alternative to circumvent the non-uniqueness problem is the Burton-Miller method \cite{Burton-Miller} which use a linear combination of CBIE and its normal derivative as the boundary integral to solve. {\color{red} emphasize the general BEM NOT BEM for acoustics}

The boundary element method (BEM) has been a most important numerical method in science and engineering. Its unique advantages includes the highly accurate solution on the boundary, the reduction of dimensionality, and the incomparable superior in solving infinite or semi-infinite field problems, etc. Historically, relatively low-order discretizations have been used with geometries modelled using first order elements and surface variables modelled to zero or first order on those elements. Recently, however, there has been increasing interest in the use of high order methods,
in order to obtain extra digits of precision with comparatively small additional effort. Successful usage are seen in acoustics \cite{nystrom}, electromagnetics \cite{ACA,James}, elasticity \cite{GD02}, aerodynamics \cite{AIAA}, to name a few.
One of the main difficulties in using high order BEM is the lack of efficient methods for accurately
evaluating various singular integrals over curved elements.
For the well-developed low order methods, there is no great difficulty since robust analytical and numerical integration schemes are generally available for planar elements \cite{constant element, fata, extraction}.
When concerned with high order elements, however, a fully numerical method is required; various techniques have been proposed, for example, the singularity subtraction \cite{Guiggiani}, special purpose quadrature \cite{rokhlin, James, machine, Carley} and the variable transformation methods \cite{wilton, wilton2, duffy, transformation}.

In \cite{Guiggiani}, Guiggiani \emph{et al} proposed a unified formula (Guiggiani's method for short) for treating various order of singular integrals on curved elements. It is a singularity subtraction method, and has found extensive use in BEM.
In this method the singular parts are extracted from the integrand and treated analytically. The remaining parts are regular so that conventional Gaussian quadrature can be employed, but the number of quadrature points needed would be large, depending on the regularity of the associated integrands.
The special purpose quadrature can be used to substantially reduce the number of quadrature points, besides it is more robust and highly accurate.
Most recently, James Bremer \cite{James} proposed such a method which can achieve machine precision.
The problem with the special purpose quadrature is that its construction somewhat complicated and time-consuming.
Variable transformation method, also known as singularity cancelation, eliminate the singularity of the integrand by the null Jacobian at the field point through a proper change of variables. Although simple to implement, it is generally hard to be used in handling hypersingular integrals.

In most existing integration methods, including those mentioned above, polar coordinates transformation (PCT) always serves as a common base \cite{Guiggiani, James, Carley, wilton, transformation, Gao}. It converses the surface integral into a double integral in radial and angular directions.
Many works have been done on dealing with the singularity in the radial direction; numerical integration on the angular direction, however, still deserves more attention. In fact, after singularity cancelation or subtraction, although the integrand may behaves very well in the radial direction, its behavior in the angular direction would be much worst, so that too many quadrature points are needed.
Especially when the field point lies close to the boundary of the element, one can clearly observe near singularity of the integrand in the angular direction. Unfortunately, this case is frequently encountered in using high order elements, especially for non-conform elements as will be demonstrated by a Nystr\"{o}m BEM in this paper. Similar problems have been considered in work about the nearly singular BEM integrals. Effective methods along this line are the subdivision method \cite{re-splitting}, the Hayami transformation \cite{Hayami}, the sigmoidal transformation \cite{sigmodial}, etc. For the singular BEM integrals, the angular transformation has been used for computing weakly singular integrals over planar element \cite{wilton}.

Another problem of the PCT is how to find a proper planar domain to establish the polar coordinates. In usual practice, the integral is carried out over standard reference domain (i.e., triangle in this paper) in intrinsic coordinates. However, the reference triangle is independent of the shape of the curved element, and thus the distortion of the element is brought into the integrand, which can cause near singularity in the angular direction. Consequently, the performance of quadratures is highly sensitive to the shape of element.
The above two problems in the angular direction were considered and resolved by special purpose quadrature rule in \cite{James}; whereas, as mentioned before, the algorithm has its own overhead.

In this paper, two strategies are proposed to overcome two problems in the angular direction, respectively.
First, a conformal transformation is carried out to map the curved physical element onto a planar integration domain. Since it is
conformal at the field point, the resultant integration domain perseveres the shape of the curved
element. Second, a new sigmoidal transformation is introduced to alleviate the near singularity caused by the closeness of the field point to the element boundary. The two proposed techniques, when combined with Guiggiani's method, lead to a unified, efficient and robust numerical integration methods for various singular integrals in high order BEM.
As a byproduct of the conformal transformation, the line integral in Guiggiani's method can be evaluated in close form.

The paper is organized as follows. The Nystr\"om BEM for acoustics and singular integrals encountered are reviewed in section \ref{BEM}. Section \ref{polar coordinates} describes the Guiggiani's unified framework for treating various order of singular integrals, with emphasis on the two reasons which render the poor performance of polar coordinates methods. Two efficient transformations which represent the main contribution of this paper are proposed in section \ref{integral}. Numerical examples are given in section \ref{examples} to validate the efficiency and accuracy of the present methods. Section \ref{conclusions} concludes the paper with some discussions.

\section{Model problem: acoustic BIEs and Nystr\"om discretization} \label{BEM}

The method presented in this paper will be used in solving the acoustic Burton-Miller BIE, which is briefly recalled here. The BIE is solved by using the Nystr\"om method with the domain boundary being partitioned into curved quadratic elements. Various singular integrals that will be treated in the following sections are summarized.

\subsection{Acoustic Burton-Miller formulation} \label{BEM:B-M}

The time harmonic acoustic waves in a homogenous and isotropic acoustic medium $\Omega$ is described by the following Helmholtz equation
\begin{equation}\label{eq:helmholtz}
    \nabla^2u(\X)+k^2u(\X)=0,\quad \forall \X \in \Omega,
\end{equation}
where, $\nabla^2$ is the Laplace operator, $u(\X)$ is the sound pressure at the point $\X=(x_1,\,x_2,\,x_3)$ in the physical coordinate system, $k=\omega/c$ is the wavenumber,
with $\omega$ being the angular frequency and $c$ being the sound speed. For static case $k=0$, \eqref{eq:helmholtz} becomes the Laplace equation.
By using Green's second theorem, the solution of Eq. \eqref{eq:helmholtz} can be expressed by integral representation
\begin{equation}\label{eq:IE}
    u(\fpt)+\optrD{u}=\optrS{q}+u^I(\fpt),\quad \forall \fpt \in \Omega,
\end{equation}
where $\X$ is a field point and $\Y$ is a source point on boundary $S$; $q(\Y)=\partial u(\Y)/\partial\un(\Y)$ is the normal gradient of sound pressure; $\un(\Y)$ denotes the unit normal vector at the source point $\Y$. The incident wave $u^I(\X)$ will not be presented for radiation problems.
The three dimensional fundamental solution $G$ is given as
\begin{equation}\label{eq:fundsol}
    G(\X,\Y)=\frac{\ue^{ikr}}{4\pi r},
\end{equation}
with $r=\norm{\X-\Y}$ denoting the distance between the source and the field points.

Before presenting the BIEs it is convenient to introduce the associated single, double, adjoint and hypersingular layer operators which are denoted by $\mathcal{S}$, $\mathcal{D}$, $\mathcal{M}$ and $\mathcal{H}$, respectively; that is,
\begin{subequations} \label{eq:operator}
\begin{align}
    \opS q(\spt)&=\optrS{q}, \label{eq:single}\\
    \opD u(\spt)&=\optrD{u}, \label{eq:double}\\
    \opM q(\spt)&=\optrM{q}, \label{eq:adjoint}\\
    \opH u(\spt)&=\optrH{u}. \label{eq:hypersingular}
\end{align}
\end{subequations}
The operator $\mathcal{S}$ is weakly singular and the integral is well-defined,
while the operators $\mathcal{D}$ and $\mathcal{M}$ are defined in Cauchy principal value sense (CPV).
The operator $\mathcal{H}$, on the other hand, is hypersingular and unbounded as a map from the space of smooth functions on $S$ to itself. It should be interpreted in the Hadamard finite part sense (HFP). Denoting a vanishing neighbourhood surrounding $\X$ by $s_{\varepsilon}$, the CPV and HFP integrals are those after extracting free terms from a limiting process to make $s_{\varepsilon}$ tends to zero in deriving BIEs \cite{Guiggiani}. %For explicit expressions of the above four kernels, see \cite{expression of kernels}.

Letting the field point $\X$ approach the boundary $S$ in Eq. \eqref{eq:IE} leads to the CBIE
\begin{equation} \label{eq:CBIE}
    C(\fpt)u(\fpt)+\mathcal{D} u(\fpt) =\mathcal{S} q(\fpt)+u^I(\fpt),\quad \fpt\in S,
\end{equation}
where, $C(\X)$ is the free term coefficient which equals to $1/2$ on smooth boundary.
By taking the normal derivative of Eq. \eqref{eq:IE} and letting the field point $\X$ go to boundary $S$,
one obtains the hypersingular BIE (HBIE)
\begin{equation} \label{eq:HBIE}
    C(\fpt)q(\fpt)+\mathcal{H} u(\fpt) =\mathcal{M} q(\fpt)+q^I(\fpt),\quad \fpt\in S.
\end{equation}
Both CBIE and HBIE can be applied to calculate the unknown boundary values of interior acoustic problems.
For an exterior problem, they have a different set of fictitious frequencies at which a unique solution can't be obtained.
However, Eqs. \eqref{eq:CBIE} and \eqref{eq:HBIE} will always have only one solution in common. Given this fact, the Burton-Miller formulation which is a linear combination of Eqs. \eqref{eq:CBIE} and \eqref{eq:HBIE} (CHBIE) should yield a unique solution for all frequencies \cite{Burton-Miller}
\begin{equation} \label{eq:Burton-Miller}
\begin{split}
    C(\fpt)u(\fpt)+\left( \opD+\alpha\opH \right)u(\spt)-u^I(\fpt)
        =\left( \opS+\alpha\opM\right)q(\spt)-\alpha\left[ c(\fpt)q(\fpt)-q^I(\fpt) \right],\quad \fpt\in S,
\end{split}
\end{equation}
where, $\alpha$ is a coupling constant that can be chosen as $i/k$.

\subsection{Nystr\"om method and the singular integrals} \label{BEM:Ny-sin}

In this paper, the Nystr\"om method is used to discretize the BIEs. Let $\mathcal{K}$ be one of the integral operators in \eqref{eq:operator} and $K$ be the associated kernel function. Divide the problem into two regions, a region near and far from the field point $\X$. If the element locates in far field, Nystr\"om method replaces the integral operator $\mathcal{K}$ with a summation under a quadrature rule
\begin{equation} \label{eq:far field}
    \int_{\Delta S} K(\X_i,\Y)u(\Y) \ud S(\Y) \cong \sum_j \omega_j K(\X_i,\Y_j) u(\Y_j), \quad \Delta S \in S\backslash D_i,
\end{equation}
where, $\fpt_i$ and $\spt_j$ are quadrature points, $D_i$ is the near field of $\fpt_i$, $\omega_j$ is the $j$th weight over element $\Delta S$.
Such quadrature rules can be obtained by mapping the Gaussian quadrature rules onto the parameterization of $\Delta S$.

If the element locates in near field, however, the kernels exhibit singularities or even hypersingularities. As a result, conventional quadratures fail to give correct results. In order to maintain high-order properties, the quadrature weights are adjusted by a local correction procedure. Thus \eqref{eq:far field} becomes
\begin{equation} \label{eq:near field}
    \int_{\Delta S} K(\X_i,\Y)u(\Y) \ud S(\Y) \cong \sum_j \bar\omega_j(\X_i) u(\Y_j), \quad \Delta S \in D_i,
\end{equation}
where $\bar\omega_j(\X_i)$ represents the modified quadrature weights for specialized rule at the singularity.
The local corrected procedure is performed by approximating the unknown quantities using linear combination of polynomial basis functions
which are defined on intrinsic coordinates (Fig.\ref{fig:element}). Modified weights are obtained by solving the linear system
\begin{equation} \label{eq:local correction}
    \sum_j \bar \omega_j \phi^{(n)}(\spt_j)=\int_{\Delta S}K(\fpt_i,\spt)\phi^{(n)}(\spt)\ud S(\spt), \quad \Delta S \in D_i,
\end{equation}
where $\phi^{(n)}$ are polynomial basis functions. For $2$th Nystr\"om method used in this paper, $\phi^{(n)}$ employed are of the form
\begin{equation}
    \phi^{(n)}(\xi_1,\xi_2)=\xi_1^p\xi_2^q, \quad p+q \le 2,
\end{equation}
where, $p$ and $q$ are integers, $\xi_1$ and $\xi_2$ denote local intrinsic coordinates.

The right hand side integrals in Eq. \eqref{eq:local correction} are of crucial importance to the accuracy of the Nystr\"om method.
They are referred to as singular integrals when $\X_i$ lies on the element, and nearly singular integrals when $\X_i$ is closed to but not on the element.
This paper deals with the integrals in the first case, the nearly singular integrals are treated via an recursive subdivision quadrature.
When concerned with the first three operators in equation \eqref{eq:operator}, the integrals have a weak singularity of $r^{-1}$,
while the other operator $\mathcal{H}$ is hypersingular of order $r^{-3}$, as $r\to0$.

\section{An unified framework for singular integrals} \label{polar coordinates}

Since various order of singularities appear in Eq. \eqref{eq:Burton-Miller} or many other BIEs, it is advantage to find a unified formula to treat these integrals in the same framework. In such a way, these integrals can be implement in just one program so that the computational cost will be reduced. By expansion of the singular integrands in polar coordinates, the three types of singular integrals considered in this paper can be handled in a unified manner by using the formula proposed by Guiggiani \cite{Guiggiani}. %The polar transformation method has been extensively used in treating singular integrals in BEM since one order of singularity can be cancelled out by the Jacobian of the transformation. The formula treating hypersingular integrals proposed by Guiggiani can be explained in a unified framework.
Despite of this advantage the method can be of low efficiency in practical usage; the very reasons are then explained.

\subsection{Polar coordinates transformation}\label{s:polar_coord}

\begin{figure}\label{fig:cur_ref}
    \centering
    \includegraphics[width = 0.5\textwidth]{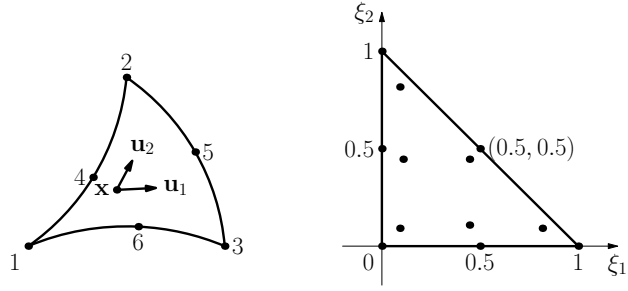}
    \caption{Description of curved triangle. Left hand image: curved triangle and a field point;
        right hand image:the reference triangle and typical field points distribution of $2$th order Nystr\"om method
        in intrinsic coordinate system} \label{fig:element}
\end{figure}

\begin{figure}[h]
    \centering
    \includegraphics[width = 0.5\textwidth]{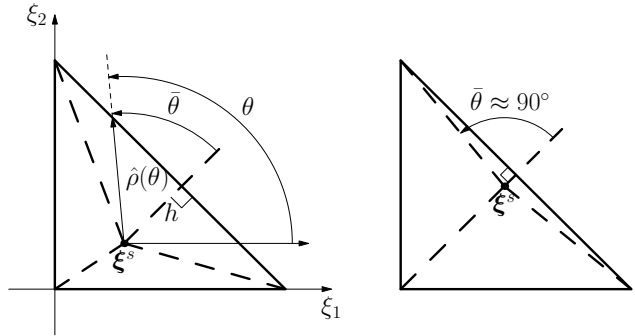}
    \caption{Integration under polar coordinates.
        Left: Polar coordinates transformation; Right: near singularity as the field point approaches the boundary} \label{fig:polar}
\end{figure}

Following a common practice in the BEM, the curved element is first mapped onto a region $\Delta$ of standard shape in the parameter plane (Fig. \ref{fig:element}). In this case, the integral must be evaluated in Eq. \eqref{eq:local correction} are of the form
\begin{equation} \label{eq:integral}
    I=\int_{\Delta} K(\X,\Y(\bm{\xi}))\phi(\bm{\xi})\norm{\J(\bm \xi)}\ud \xi_1 \ud \xi_2,
\end{equation}
where $\norm{\J}$ is the transformation Jacobian from global coordinates to the intrinsic coordinates,
\begin{equation*}
    \J = \left[\pfrac{\spt}{\xi_1} \quad \pfrac{\spt}{\xi_2} \right], \qquad \norm{\J(\bm \xi)}=\norm{\pfrac{\spt}{\xi_1} \times \pfrac{\spt}{\xi_2}}.
\end{equation*}
Then,
polar coordinates $(\rho,\theta)$ centered at $\bm{\xi}^{s}$ (the image of $\X$ on intrinsic plane) are defined in the parameter space (Fig. \ref{fig:polar})
\begin{equation*}
\left\{
    \begin{split}
        \xi_1=\xi_1^{(s)}+\rho\cos\theta\\
        \xi_2=\xi_2^{(s)}+\rho\sin\theta
    \end{split}
\right.
\end{equation*}
so that $\ud\xi_1\ud\xi_2=\rho\ud\rho\ud\theta$. Due to piecewise smooth property of the boundary of $\Delta$,
the triangle is split into three sub-triangles. The associated integral of Eq. \eqref{eq:integral} now becomes
\begin{equation} \label{eq:polar}
    I=\lim_{\rho(\varepsilon) \to 0} \sum_{j=1}^3 \int_{\theta_{j-1}}^{\theta_j}\int_{\rho(\varepsilon)}^{\hat{\rho}(\theta)}
     K(\rho,\theta)\phi(\rho,\theta)\abs{\J(\rho,\theta)} \rho\ud \rho \ud\theta,
\end{equation}
where $\hat{\rho}(\theta)$ gives a parametrization of the boundary of $\Delta$ in polar coordinates,
$(\theta_{j-1},\theta_j)$ are three intervals on which $\hat{\rho}(\theta)$ is smooth. The limiting process means CPV or HFP integral mentioned before, although for well-defined integral operator $\mathcal{S}$, the limiting is not necessary.
According to the geometrical relationship in Fig.\ref{fig:polar}
\begin{equation} \label{eq:boundary}
    \hat{\rho} (\theta)=\frac{h_{j}}{\cos\bar{\theta}}.
\end{equation}
where, $h_j$ is the perpendicular distance from $\bm\xi^s$ to $j$th side of the planar triangle and $\bar{\theta}$ is the angle from the perpendicular to a point $\bm\xi$ (Fig.\ref{fig:polar}). In each sub-triangle, $\bar\theta$ equals to $\theta$ minus a constant.

For singularity of order no more than 3, the integrand of \eqref{eq:polar}, denoted by $F(\rho,\theta)$, can be expressed as series expansion under polar coordinate (\cite{kernel,Guiggiani})
\begin{equation} \label{eq:expansion}
    F(\rho,\theta)=K(\rho,\theta)\phi(\rho,\theta)\abs{\J} \rho\ud \rho=\frac{f_{-2}(\theta)}{\rho^2}+\frac{f_{-1}(\theta)}{\rho}+f_{0}(\theta)+\rho f_1(\theta)+\rho^2 f_2(\theta)+\dots=\sum^{\infty}_{i=p} \rho^i f_i(\theta),
\end{equation}
where, $f_i$ are just functions of $\theta$, integer $p$ is determined by the order of singularity,
\begin{equation*}
    p = \left\{
          \begin{array}{ll}
            0, & \hbox{weakly singular;} \\
            -1, & \hbox{strongly singular;} \\
            -2, & \hbox{hyper-singular.}
          \end{array}
        \right.
\end{equation*}

First, let's consider the hypersingular integrals. Due to the appearance of the two terms with $(i=-2,\,-1)$, the integration of $F(\rho,\theta)$ must be performed in the HFP sense; for more details see Guiggiani's work \cite{Guiggiani}. The resultant formula for hypersingular integrals is given by
\begin{equation} \label{eq:Guiggiani}
    \begin{split}
        I=I_1+I_2=& \underbrace{    \sum_{j=1}^3 \int_{\theta_{j-1}}^{\theta_j} \int_{0}^{\hat{\rho}(\theta)}\left[ F(\rho,\theta)-\left(\frac{f_{-2}(\theta)}{\rho^{2}}+\frac{f_{-1}(\theta)}{\rho}\right) \right]\ud \rho \ud \theta }_{I_1}\\
         &+ \underbrace{ \sum_{j=1}^3 \int_{\theta_{j-1}}^{\theta_j}\left( f_{-1}(\theta)\ln\hat{\rho}(\theta)-f_{-2}(\theta)\frac{1}{\hat{\rho}(\theta)} \right)\ud \theta }_{I_2}.
    \end{split}
\end{equation}

The strongly and weak singular integrals can be treated in a similar manner based on expansion \eqref{eq:expansion}. The resultant computing formulas can also be written in the above form except that for strongly singular integrals $f_{-2}=0$, and for weak singular integrals $f_{-2}=f_{-1}=0$ (thus $I_2 =0$). Therefore, formula \eqref{eq:Guiggiani} actually provide an unified approach to evaluate singular integrals in BEM.

In \eqref{eq:Guiggiani} the double integral $I_1$ and single integral $I_2$ are all regular, thus one may conclude that it is sufficient to guarantee numerical accuracy in evaluating these two integrals, it would be nevertheless too expensive in practical usage, especially when used in high order Nystr\"om method considered in this paper. The difficulties and the corresponding solutions will be presented in the following sections.

\subsection{Difficulties in evaluating $I_1$} \label{properties}

It is the computation of $I_1$ that accounts for the main cost for evaluating the various singular integrals.
The integrand of $I_1$ can be approximated by polynomials in $\rho$; that is,
\begin{equation} \label{eq:I0}
    I_1 =   \sum_{j=1}^3 \int_{\theta_{j-1}}^{\theta_j}\int_{0}^{\hat{\rho}(\theta)}\left( f_{0}(\theta)+\rho f_1(\theta)+\rho^2 f_2(\theta)+\dots \right)\ud \rho \ud \theta.
\end{equation}
The number of terms in this approximation is determined by the kernel function,
the order of basis function and the flatness of the associated element. The relative flatness of element is a basic requirement in BEM in order to guarantee the accuracy. Consequently, it appears that the integrand of $I_1$ can be well approximated by low order polynomials in $\rho$ in solving many problems including Laplace, Helmholtz, elasticity and so forth, using quadratic elements. This implies that low order Gaussian quadratures are sufficient for numerical integration in $\rho$.

In angular $\theta$ direction, however, two difficulties are frequently encountered which severely retard the convergence rate of Gaussian quadratures. To show this, preforming integration in $\rho$ in \eqref{eq:I0} one obtains
\begin{equation} \label{eq:nearly}
    I_1=\sum_{j=1}^3\int_{\theta_{j-1}}^{\theta_j}\left(\hat{\rho}(\theta)f_{0}(\theta)+\frac{1}{2}\hat{\rho}^2(\theta)f_1(\theta)+\cdots\right)\ud\theta.
\end{equation}
The difficulties in computation of $I_1$ are caused by the near singularities of $f_i(\theta)$ and $\hat\rho(\theta)$.

\subsubsection{Near singularity in $f_i(\theta)$} \label{s:fi}

Functions $f_i(\theta)$ can be expressed as (\ref{App:coefficient})%\cite{kernel}
\begin{equation}\label{eq:fi}
    f_i(\theta)=\frac{\tilde{f}_i(\theta)}{A^{\alpha}(\theta)},
\end{equation}
where $\tilde{f}_i$ are regular trigonometric functions and $\alpha$ is integer determined by the subscript $i$. Function $A(\theta)$ is depend on the shape of the element and parametric coordinate system (see \cite{Guiggiani} for a definition of $A(\theta)$). Specifically, let $\U_1$, $\U_2$ be the two column vectors of the Jacobian matrix $\J$ which spans the space tangent to the element at the point $\X$, i.e.
\begin{equation}
    \U_1=\left.\pfrac{\spt}{\xi_1}\right|_{\Y=\X} \quad \text{and} \quad \U_2=\left.\pfrac{\spt}{\xi_2}\right|_{\Y=\X}.
\end{equation}
Then $A(\theta)$ is given by
\begin{equation}\label{Atheta}
\begin{split}
    A(\theta)=&\sqrt{\norm{\U_1}^2\cos^2\theta+\U_1\cdot\U_2\sin2\theta+\norm{\U_2}^2\sin^2\theta}\\
             =&\sqrt{\U_1\cdot\U_2\sin2\theta+\frac{1}{2}\left(\abs{\U_1}^2-\abs{\U_2}^2\right)\cos2\theta
             +\frac{1}{2}\left(\abs{\U_1}^2+\abs{\U_2}^2\right)}\\
             =& \sqrt{\frac{1}{2}\left(\abs{\U_1}^2+\abs{\U_2}^2\right)\left[\mu\sin(2\theta+\varphi)+1\right]}.
\end{split}
\end{equation}
If we let $\lambda=\abs{\U_1}/\abs{\U_2}$ and $\cos\gamma=\frac{\U_1\cdot\U_2}{\abs{\U_1}\abs{\U_2}}$, then
\begin{equation*}
    \varphi=\arctan\frac{\lambda^2-1}{2\lambda\cos\gamma},
\end{equation*}
and
\begin{equation} \label{eq:l}
    \mu=\sqrt{1-\frac{4\sin^2\gamma}{(\lambda+\lambda^{-1})^2}}<1.
\end{equation}

It can be seen from \eqref{Atheta} that, if $\mu \rightarrow 1$, there exist two points $\theta \in [0,2\pi]$ such that $A(\theta)\rightarrow 0$. Thus function $f_i$ tends to be nearly singular. The circumstance $\mu \rightarrow 1$ occurs in two cases according to Eq. \eqref{eq:l}: (1) the aspect ratio of the element is large, i.e., $\lambda$ approaches $0$ or $\infty$; (2) peak or big obtuse corner appear in the element which lead to $\sin\gamma\to0$.
Both of these two cases indicate a distorted shape of the element.

\begin{figure}[htb]
    \centering
    \includegraphics[width = 0.55\textwidth]{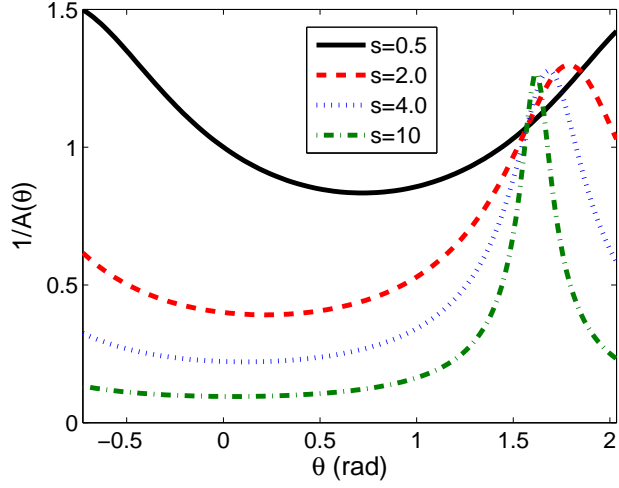}
    \caption{The plot of $1/A(\theta)$ for various value of $s$.
    The aspect ratio of the associated element increased with $s$.
    The curved element is described in section \ref{example1}, with field point $(b)$} \label{fig:A}
\end{figure}

To illustrate the influence of the element aspect ratio on the behavior of function $1/A(\theta)$ which represents the smoothness of $f_i$, consider the curved element in figure \ref{fig:ex1}; see section \ref{example1} for more detailed descriptions. The aspect ratio of the element is controlled by $s$; larger $s$ implies more distorted shape. Let $b$ be the singular point. The plots of $1/A(\theta)$ in the interval corresponding to the sub-triangle (2-b-3) in figure \ref{fig:ex1} with various $s$ are exhibited in Fig. \ref{fig:A}. It is clear that $1/A(\theta)$ varies more acutely as the increase of the aspect ratio.

The above analysis shows how element shape affects $f_i$ and thus the integrand in \eqref{eq:nearly}. In an intuitionistic manner,
the near singularity in $f_i$ is induced by the fact that the integration plane $\Delta$ in intrinsic coordinate system is independent on the shape of element, as a result the distortion is brought into the integrand.
One can suppose that if the integration is performed over another planar triangle which reflects the distortion of the element,
the near singularity should be eliminated. This is the main idea of the our new method in section \ref{change}.

\subsubsection{Near singularity in $\hat \rho(\theta)$}\label{s:rho}

In addition to the near singularity in $f_i$,
another obstacle retards the convergence of the numerical quadrature for \eqref{eq:nearly} is the near singularity in $\hat{\rho}(\theta)$, which can be clearly seen from \eqref{eq:boundary}. When the field point $\X$ lies near to the boundary of the element, the restrict of $\bar\theta$ approaches $\pm\pi/2$, thus
the denominator $\cos\bar{\theta}$ is close to $0$ at the two ends of the interval $(\theta_{j-1},\theta_j)$ (see Fig.\ref{fig:polar}).
The effect of the near singularity in $\hat{\rho}(\theta)$ on the total behavior of the integrand $F(\rho, \theta)$ is demonstrated by the left plots of Fig. \ref{fig:single} (a) and (b), where the integrand have large peaks near the two ends of the interval.

Unfortunately, the situation that causes the near singularity in $\hat{\rho}(\theta)$ is ubiquitous in using high order Nystr\"om method. See Fig. \ref{fig:cur_ref} for a typical distribution of field points in $2$th order Nystr\"om method.

\section{Efficient transformation methods} \label{integral}

How to effectively resolve the above mentioned two difficulties is crucial to the accurate evaluation of the BEM singular integrals.
More recently, special quadratures are constructed for this purpose in \cite{James}.
Although accurate and robust numerical results are reported, it is noticed that the abscissas and weights of the special quadratures
are depend on the singular (field) point as well as the element on which the integral is defined. The construction of the quadrature for integral can be rather complicated and time-consuming.

In this section, however, we propose a more simple yet efficient method.
First, the intrinsic coordinates $(\xi_1, \xi_2)$ is transformed onto a new system which results in constant $A(\theta)$.
An additional benefit of constant $A(\theta)$ is that the line integral $I_2$ can be evaluated in closed form; see section \ref{analytic}.
Then the sigmoidal transformation is introduced to alleviate the near singularity caused by $\hat{\rho}(\theta)$.

\subsection{Conformal transformation} \label{change}

First, the near singularity caused by $A(\theta)$ in $f_i$ is considered and resolved. The idea is to introduce a new transformation under which
$A(\theta)$ becomes constant. It can be seen from \eqref{Atheta} that if
\begin{equation} \label{eq:const_condition}
    \U_1\cdot\U_2=0 \quad \text{and} \quad \norm{\U_1}=\norm{\U_2},
\end{equation}
then $A(\theta)$ will be constant, i.e. $A(\theta)=\norm{\U_1}=\norm{\U_2}$.

Relations \eqref{eq:const_condition} indicate a conformal mapping from the element to the plane at field point $\X$,
i.e., both angle and the shape of the infinitesimal neighborhood at $\X$ are preserved.
However, this condition generally can not be satisfied under intrinsic coordinates. In \cite{Hayami} Hayami proposed to connect the three corners of the curved element to establish a planar triangle which preserves the shape of the element. This operation, nevertheless, can not satisfy condition \eqref{eq:const_condition} exactly.

\begin{figure}
    \centering
    \includegraphics[width = 0.45\textwidth]{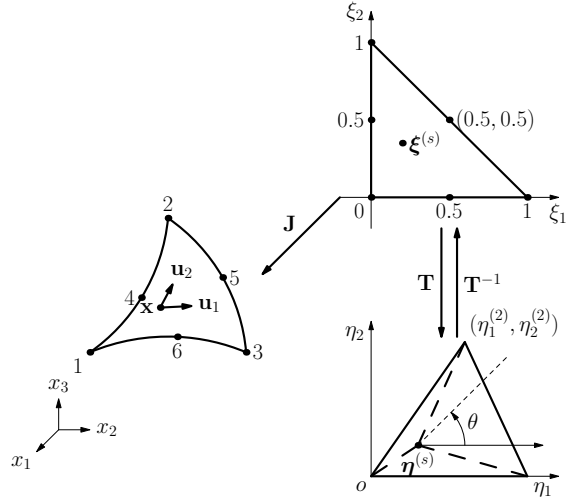}
    \caption{Transformations between coordinate systems.
        Left: curved element in global coordinates; Top right: intrinsic coordinates and the reference element;
        Bottom right: the parametric plane to perform integral} \label{fig:coordinates}
\end{figure}

Here, we propose a transformation in which the curved physical element is mapped to a triangle $\bar{\Delta}$ in plane $(\eta_1, \, \eta_2)$ as shown in Fig. \ref{fig:coordinates}. The coordinates of the three corners of $\bar{\Delta}$ are $(0,0)$, $(\eta_1^{(2)},\eta_2^{(2)})$ and $(1,0)$, with $\eta_2^{(2)}>0$. The transformation from system $(\xi_1,\,\xi_2)$ to $(\eta_1,\,\eta_2)$ can be realized by linear interpolation
\begin{equation} \label{xi_to_eta}
    \bm{\eta}=\sum_{i=1}^3 \bar{\phi}_i(\xi_1,\xi_2)\bm{\eta}^{(i)},
\end{equation}
where, $\bm{\eta}^{(i)}$ are coordinates of the three corners of $\bar{\Delta}$, $\bar{\phi}_i$ are linear interpolating functions
\begin{equation}\label{linear_basis}
\begin{aligned}
    \bar{\phi}_1&=1-\xi_1-\xi_2,\\
    \bar{\phi}_2&=\xi_2,\\
    \bar{\phi}_3&=\xi_1.
\end{aligned}
\end{equation}
Plugging \eqref{linear_basis} into \eqref{xi_to_eta} yields
\begin{equation} \label{xi_eta}
    \bm{\eta}=\mathbf{T}\bm{\xi},
\end{equation}
where, $\mathbf{T}$ is the transformation matrix
\begin{equation} \label{eq:T}
    \mathbf{T}=\begin{bmatrix}
                         1 & \eta_1^{(2)}\\
                         0 & \eta_2^{(2)}
               \end{bmatrix} \quad \text{and} \quad
\mathbf{T}^{-1}=\begin{bmatrix}
                        1 & -\frac{\eta_1^{(2)}}{\eta_2^{(2)}} \\
                        0 &\frac{1}{\eta_2^{(2)}}
                       \end{bmatrix}.
\end{equation}

The Jacobian matrix at $\X$ from physical coordinates to $(\eta_1, \, \eta_2)$ coordinates, denoted by $[ \bar{\U}_1\, \bar{\U}_2 ]$, can be written as
\begin{equation} \label{eq:new_jcb}
    \begin{bmatrix} \bar{\U}_1 & \bar{\U}_2 \end{bmatrix} = \begin{bmatrix} \U_1 & \U_2 \end{bmatrix} \mathbf{T}^{-1}=\begin{bmatrix} \U_1 & \frac{-\eta_1^{(2)}\U_1}{\eta_2^{(2)}}+\frac{\U_2}{\eta_2^{(2)}} \end{bmatrix}.
\end{equation}
In order to obtain constant $A(\theta)$, $\bar\U_1$ and $\bar\U_2$ have to satisfy condition \eqref{eq:const_condition}, thus one has
\begin{equation}
    \eta_1^{(2)}=\frac{\cos\gamma}{\lambda}, \quad \eta_2^{(2)}=\frac{\sin\gamma}{\lambda},
\end{equation}
where, $\lambda$ and $\gamma$ are defined in section \ref{properties}.

By using relation \eqref{xi_eta} the integral \eqref{eq:integral} can be transformed onto $(\eta_1,\,\eta_2)$ plane. One can then employ the polar coordinate transform in Section \ref{s:polar_coord}. The origin of the polar system is set as the image of $\X$ on $(\eta_1,\,\eta_2)$ plane. Then integral $I_1$ becomes
\begin{equation}
I_1=\sum_{j=1}^3 \int_{\theta_{j-1}}^{\theta_j} \int_{0}^{\hat{\rho}(\theta)}\left[ F(\rho,\theta)-\left(\frac{f_{-2}(\theta)}{\rho^{2}}+\frac{f_{-1}(\theta)}{\rho}\right) \right]  \norm{\mathbf{T}^{-1}} \rho\ud \rho \ud\theta,
\end{equation}
in which the near singularity caused by $A(\theta)$ has been successfully removed.

Numerical results show that the above transformation can always improve the numerical integration regardless the shapes of the elements being regular or irregular.

\begin{figure}
    \centering
    \includegraphics[width = 0.4\textwidth]{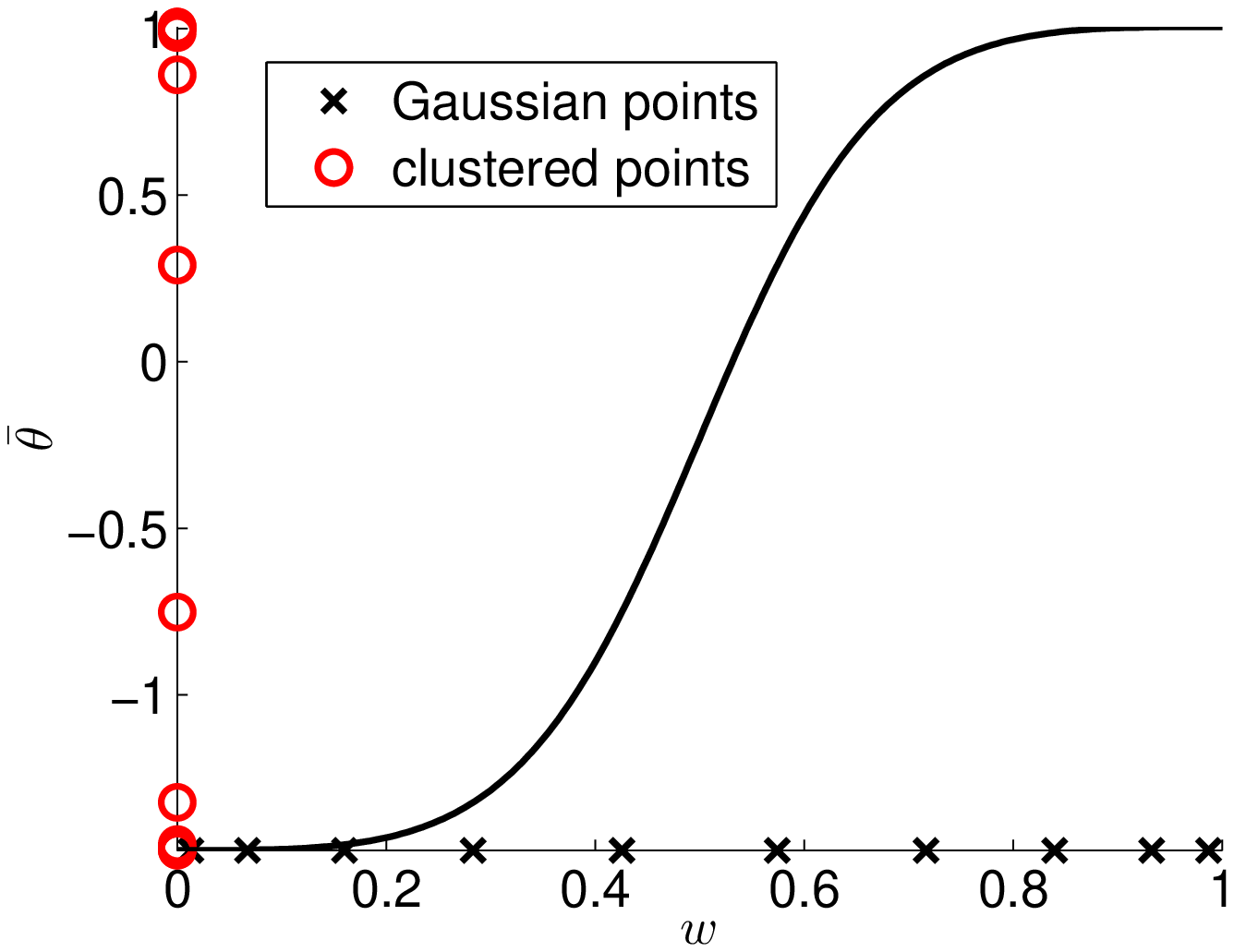}
    \includegraphics[width = 0.4\textwidth]{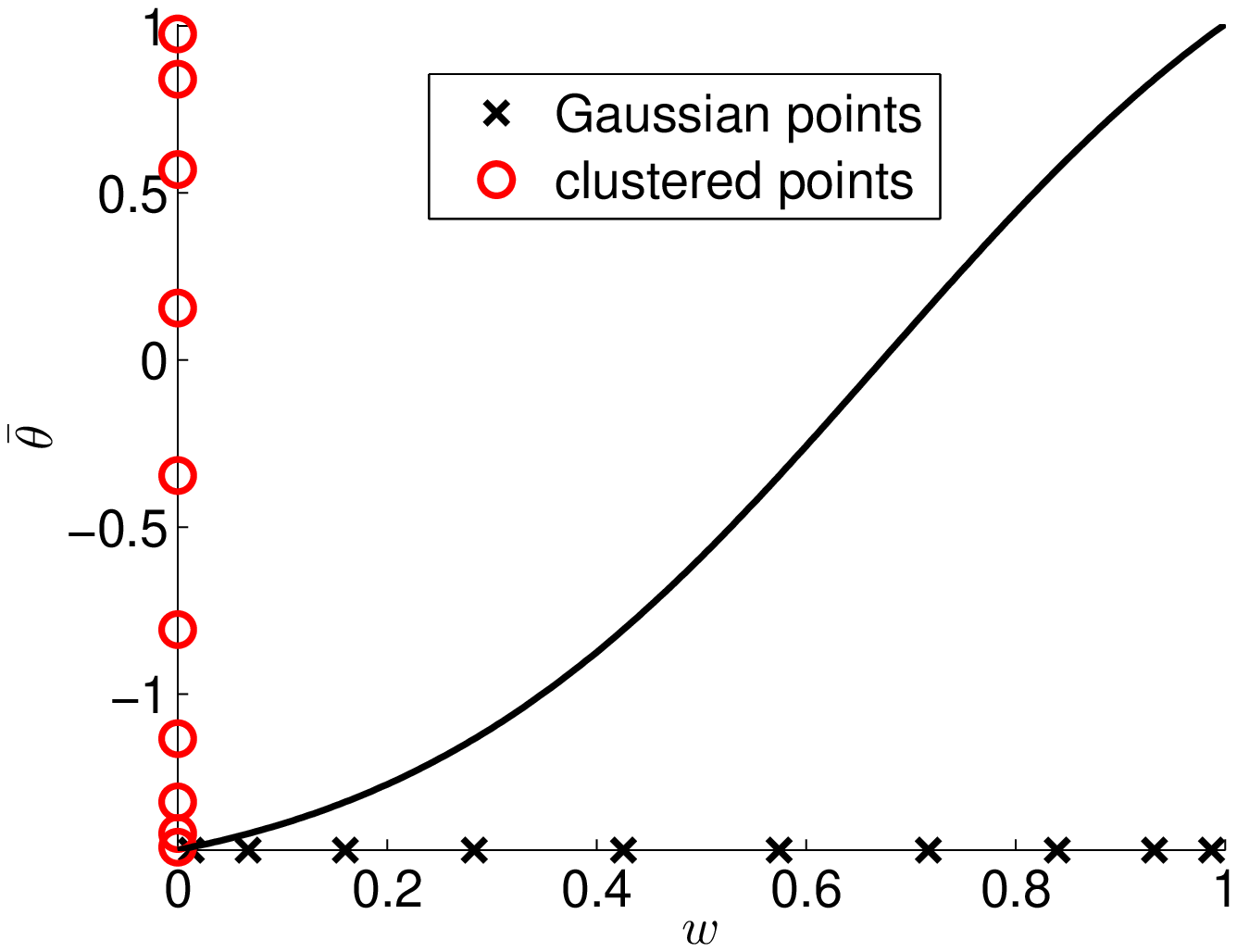}
    \caption{Sigmoidal transformation with $m=3$ to cluster 10 Gaussian quadrature points (x-axis) to the endpoints of the interval. The sub-triangle (2-b-3) in Fig. \ref{fig:ex1} is considered, whose $\bar\theta \in [-1.466,1.005]$.
        Left: A naive use of sigmoidal transformation;
        Right: Sigmoidal transformation in this paper.} \label{fig:sigmoidal}
\end{figure}

\subsection{Sigmoidal transformation} \label{S:sig-trans}

When the field point $\X$ approaches the element boundary, function $\hat{\rho}(\theta)$ and thus the integral $I_1$ becomes nearly singular.
Here a sigmoidal transformation is introduced to alleviate this problem.
The sigmoidal transformation was first proposed to calculate the singular and nearly singular integrals
in two dimensional BEM \cite{sigmodial_o}. Recently, this approach was adopted as an angular transformation
in dealing with nearly singular integrals in 3D BEM \cite{sigmodial}. One should notice that the angular transformation proposed by Khayat and Wilton \cite{wilton} can be alternatively used, but our numerical experience indicates a better overall performance by using sigmoidal transformation, especially in hypersingular case.

A sigmoidal transformation can be thought of as a mapping of the interval $[0,1]$ onto itself whose graph is {\it S}-shaped. It has the effect of translating a grid of evenly spaced points on $[0,1]$
onto a non-uniform grid with the node points clustered at the endpoints. A typical sigmoidal transformation is given by \cite{sigmodial_o}
\begin{equation} \label{eq:sigmoidal}
    \sigma(w)=\frac{w^m}{w^m+(1-w)^m}, \quad \sigma, \,w \in [0,1],\, m \ge 1.
\end{equation}
Consider one sub-triangle in figure \ref{fig:polar} in which $\bar\theta \in [\bar\theta_{j-1}, \bar\theta_{j}],\, j=1,\,2,\,3$. A naive use of the above transformation can be (see Fig. \ref{fig:sigmoidal})
\begin{equation} \label{eq:naive}
   \frac{\bar\theta-\bar\theta_{j-1}}{\bar\theta_j-\bar\theta_{j-1}} = \sigma(w).
\end{equation}
However, this would be of low efficiency. Since the integral $I_1$ tends to be nearly singular (due to $\hat\rho(\theta)$) only when $\bar\theta_{j-1}$ or $\bar\theta_{j}$ is close to right angle or both, it is thus more reasonable to cluster the quadrature nodes according to the discrepancy of $\bar\theta_{j-1}$ and $\bar\theta_{j}$ to right angles, respectively.
In addition there are cases where both
$\bar\theta_{j-1}$ and $\bar\theta_{j}$ are not very close to right angles and thus the near singularity is not severe. For these cases, one can use the Gauss quadratures directly without any transformation. Nevertheless, numerical examples in this paper show that the modification put forward below can always achieve more accurate results.

The transformation used in this paper is based on the fact that the singularity occurs at $\bar\theta=\pm\pi/2$ and is given by
\begin{equation}\label{eq:sigma}
    \frac{1}{\pi}\left(\bar{\theta}+\frac{\pi}{2}\right)= \sigma(z), \quad z = (z_{j} - z_{j-1})w + z_{j-1}, \quad w \in [0,1],
\end{equation}
where,  $z(\bar\theta_{j-1})$ and $z(\bar\theta_j)$ are the values of $z$ corresponding to $\bar\theta_{j-1}$ and $\bar\theta_{j}$ which can be easily obtained by \eqref{eq:sigmoidal}. Obviously, both transformation \eqref{eq:sigma} and \eqref{eq:naive} project $[\bar\theta_{j-1}, \bar\theta_{j}]$ onto $[0,1]$, but \eqref{eq:sigma} can cluster the quadrature points adaptively according to closeness of $\bar\theta$ to right angle, which is shown in Fig. \ref{fig:sigmoidal}. The left hand image demonstrate the part of sigmoidal transformation used in dealing with the sub-triangle (2-b-3) in Fig. \ref{fig:ex1}. Compared with full sigmoidal transformation shown left, the distribution of quadrature points are more reasonable since they are more clustered on $\bar\theta_{j-1}$ which is more closely to right angle. Besides, the points are not too away from the center. A comparison between the original sigmoidal transformation and the modification is illustrated in table \ref{tab:sig}. Without any doubt, the modified transformation is more powerful when dealing with the integrals in this paper.

The degree of cluster is impacted by parameter $m$, see \cite{sigmodial_o}. Although numerical examples indicates not too much difference in number of quadrature points when $m$ takes the value between $2$ and $3$, it should be point out that the optimum value of $m$ is $3$ for weakly singular case and $2$ for hypesingular case.
%\red{say: m impact the ...., see ref[?]. In this paper we use m=???}

Finally, the integral has the formula
\begin{equation}
    I_1=\sum_{j=1}^3 \int_{0}^{1} \int_{0}^{\hat{\rho}(\theta)}\left[ F(\rho,w)-\left(\frac{f_{-2}(w)}{\rho^{2}}+\frac{f_{-1}(w)}{\rho}\right) \right] \bar{\theta}_j'(w) \norm{\mathbf{T}^{-1}} \rho\ud \rho \ud w,
\end{equation}
where, $\bar{\theta}_j(w)$ denotes transformation \eqref{eq:sigma} on interval $[\bar\theta_{j-1}, \bar\theta_{j}]$.

\begin{table}
\centering
\caption{Comparison of relative errors using different methods in $k=0$ case. The curved element is show in Fig. \ref{fig:ex1}, with field point (b). ``n'' denotes the number of quadrature points used in angular direction.}\label{tab:sig}
\begin{tabular}[c]{c|ccc|ccc}
    \hline
                       & \multicolumn{3}{c}{single}               &\multicolumn{3}{c}{hyper}\\
    \hline
    n   & Guiggiani & Eq. \eqref{eq:naive} & Present & Guiggiani & Eq. \eqref{eq:naive}   & Present\\
    \hline
    5   & 6.24e-4 & 3.19e-3 & 9.83e-5 & 1.80e-03  & 1.57e-04                & 4.09e-05\\
    8   & 1.02e-3 & 1.16e-5 & 6.44e-7 & 1.69e-04  & 1.87e-06                & 1.49e-07\\
    10  & 3.20e-4 & 3.06e-6 & 1.26e-8 & 2.39e-05  & 5.22e-07                & 3.71e-09\\
    \hline
\end{tabular}
\end{table}

\begin{figure}
    \centering
    \subfigure[Laplace single kernel, $m=3$]{
    \includegraphics[width = 0.4\textwidth]{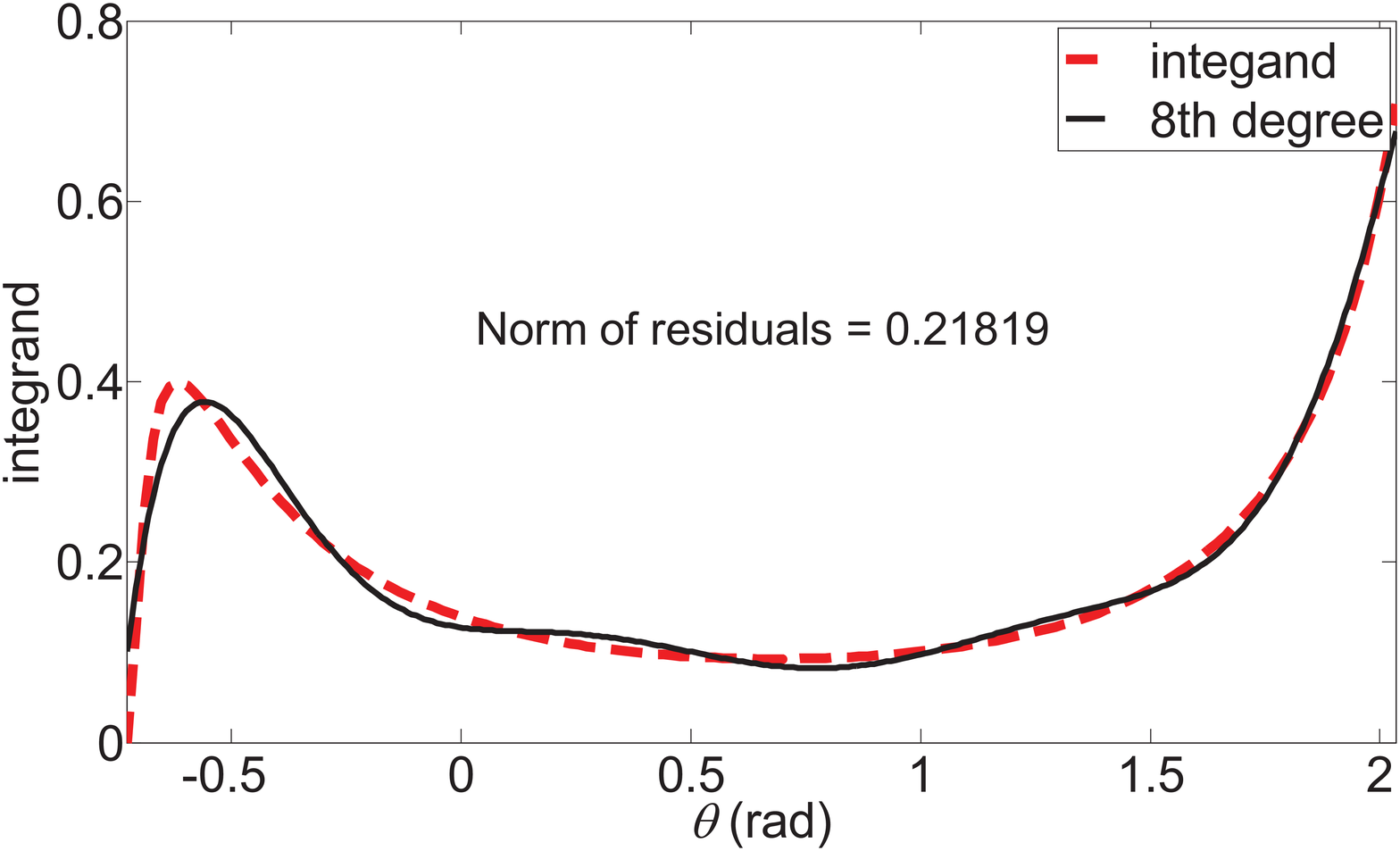}
    \includegraphics[width = 0.4\textwidth]{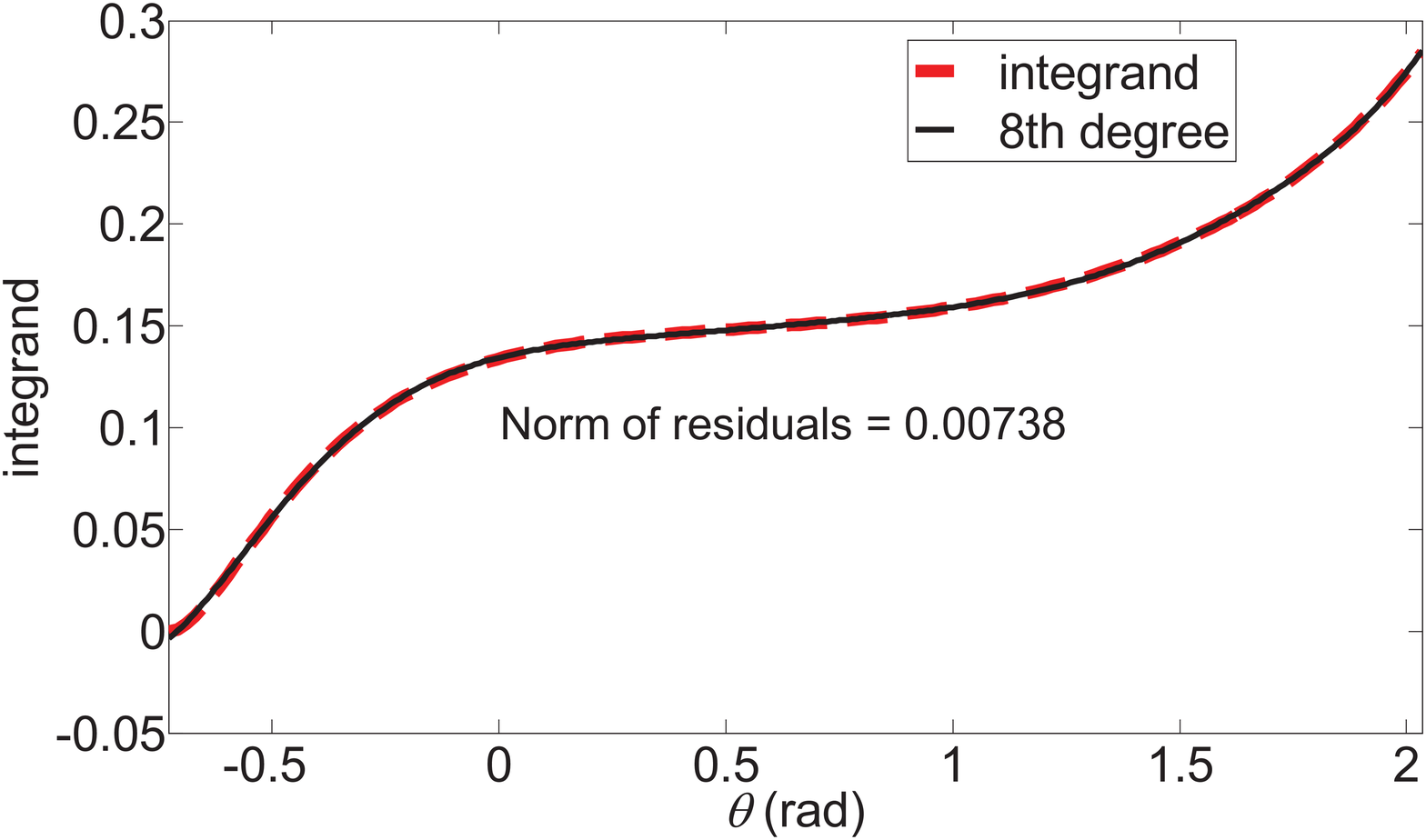}
    }\\
    \subfigure[Laplace hypersingular kernel, $m=2$] {
    \includegraphics[width = 0.4\textwidth]{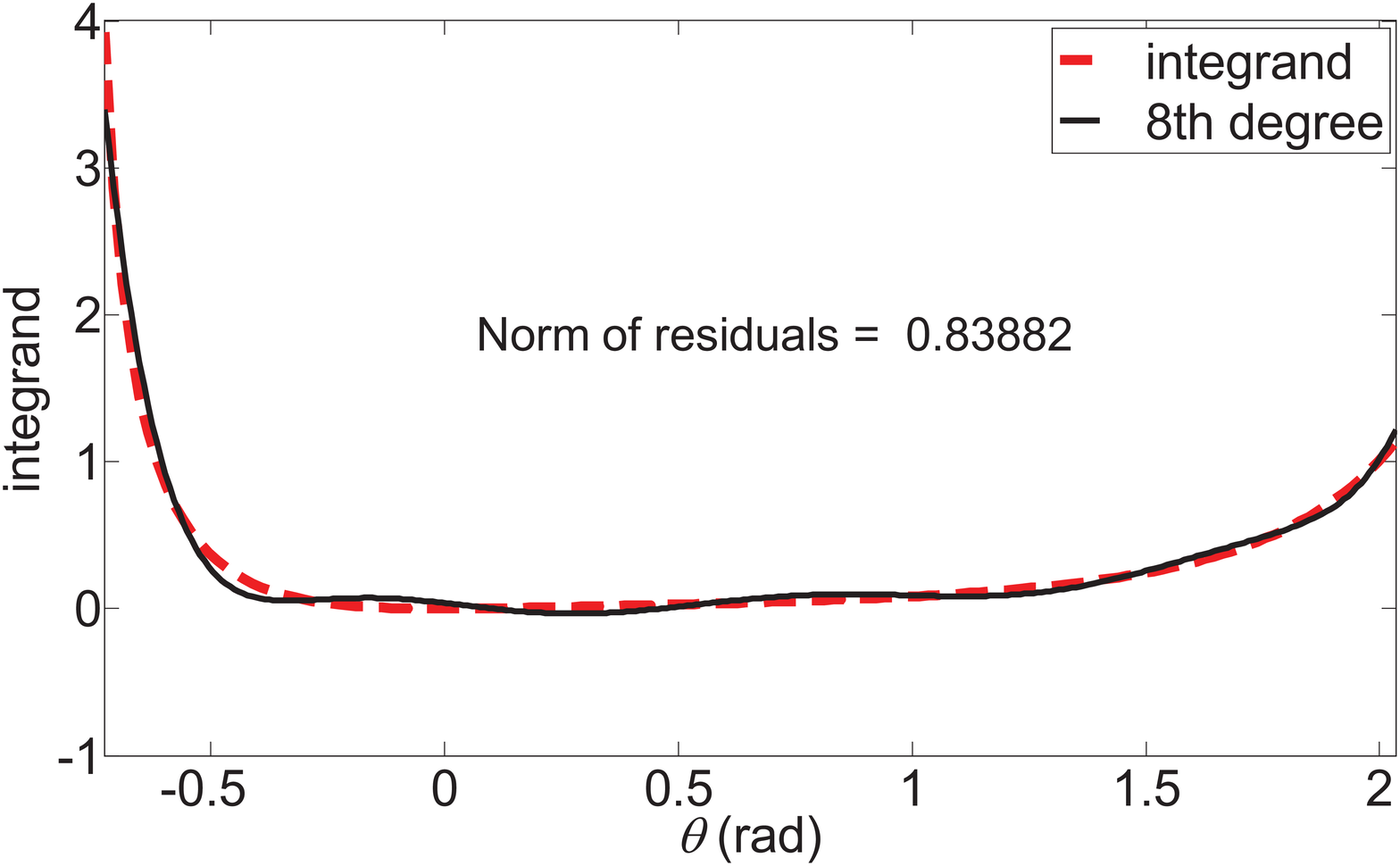}
    \includegraphics[width = 0.4\textwidth]{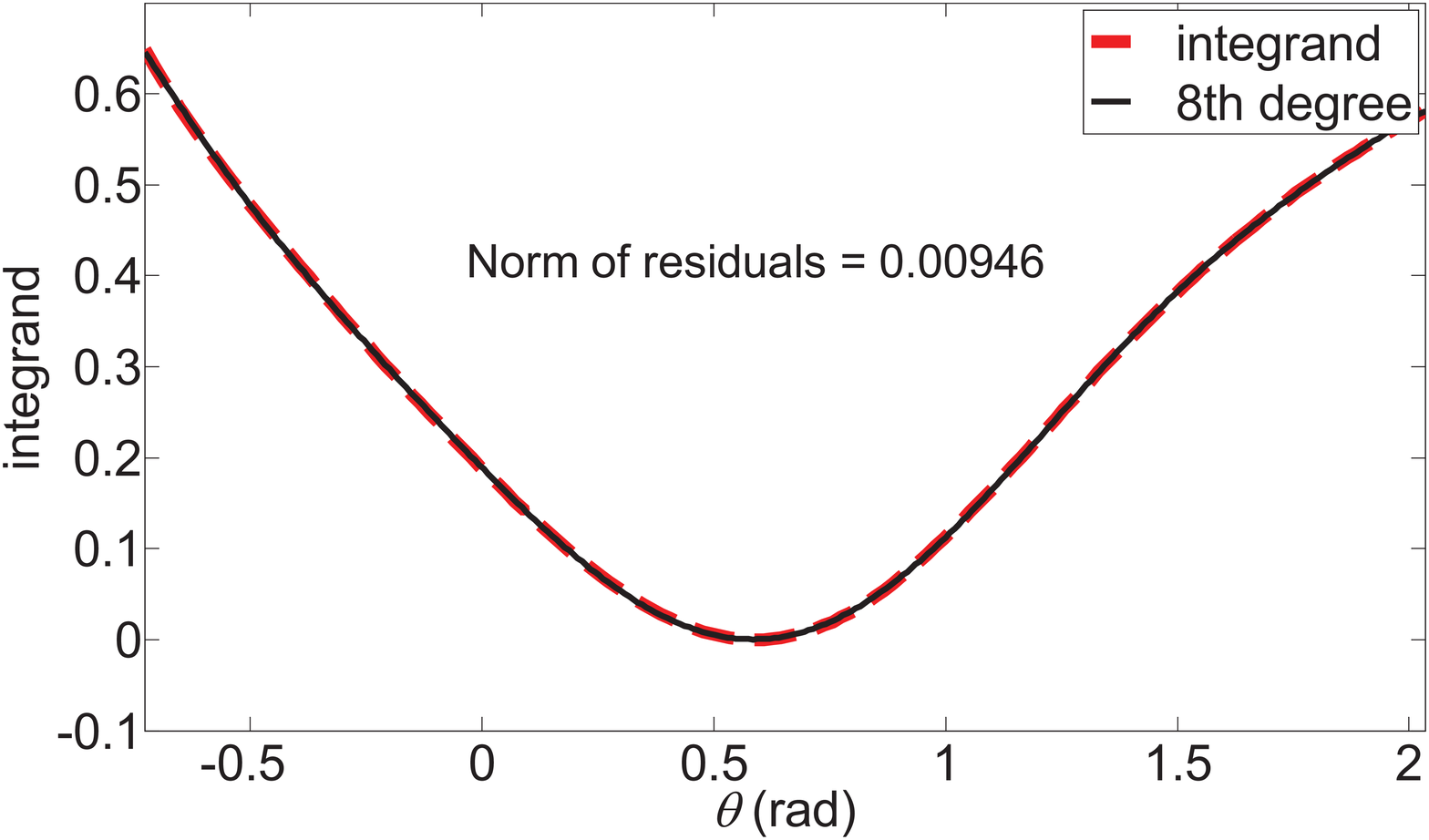}
    }
    \caption{Integrand of $I_1$ in angular direction corresponding to sub-triangle (2-b-3) in Fig. \ref{fig:ex1};the basis function is selected as $\xi_2^2$. $8$th order polynomial is plotted to fit the image.
        Left: Image of the original Guiggiani's method; Right: Image after transformations proposed in this paper.
        } \label{fig:single}
\end{figure}

The effects of the two transformations in sub-sections \ref{change} and \ref{S:sig-trans} are demonstrated in Fig. \ref{fig:single}. It can be seen that after the transformations, the integrands become more regular and can be well approximated by polynomials of order 8, which means that low order Gaussian quadratures can achieve accurate results.

\subsection{Analytical integration of the one-dimensional integrals} \label{analytic}

Consider the evaluation of the regular one dimensional integral $I_2$ in Eq. \eqref{eq:Guiggiani}, i.e., $I_2 = \sum_{j=1}^3 I^{(j)}_2$ and
\begin{equation}\label{eq:I2j}
I^{(j)}_2 =  \int_{\theta_{j-1}}^{\theta_j}\left( f_{-1}(\theta)\ln\hat{\rho}(\theta)-f_{-2}(\theta)\frac{1}{\hat{\rho}(\theta)} \right)\ud \theta.
\end{equation}

One notes that the integrand of $I_2$ suffers from the same difficulties as that of the double integral $I_1$ as stated in section \ref{properties}. The above proposed transformations can also be used to promote the efficiency of the numerical quadratures.

In this section close form expression for $I_2$ is derived. It is attributed to the fact that after the conformal transformation in section \ref{change} the denominators of $f_i$, i.e. $A(\theta)$, become constant. For Burton-Miller equation, it is readily to see from \eqref{eq:A-f1} that $f_{-2}(\theta)$ is constant on each sub-triangle thanks to the constant $A(\theta)$, and that
$f_{-1}(\theta)$ is typically made up of combinations of elementary
trigonometric functions of $\theta$ \cite{Guiggiani}
\begin{equation}\label{eq:f1}
    \begin{split}
    f_{-1}(\theta)=& c_1\cos^3\theta+c_2\cos^2\theta\sin\theta+c_3\cos\theta\sin^2\theta+c_4\sin^3\theta\\
                    &+d_1\cos\theta+d_2\sin\theta.
    \end{split}
\end{equation}
The coefficients $c_i$ and $d_i$ are given in \eqref{eq:aibi}. Here we notice that the expressions in \eqref{eq:aibi} are also applicable for Laplace equations. For other problems, such as elasticity, Stokes flow, the expressions of $f_{-2}$ and $f_{-1}$ are more complicated and can be derived analogously as in \ref{App:coefficient}.

To further simplify the derivation, the coordinate system $(\eta_1,\eta_2)$ is rotated so that
the axis $o\eta_1$ is parallel to the perpendicular line of the side of $\bar\Delta$.
Thus $\theta=\bar{\theta}$ over each sub-triangle. By substituting $\hat{\rho}(\theta)$, $f_{-2}$ and $f_{-1}$ into \eqref{eq:I2j}, the integral with $f_{-2}$ becomes
\begin{equation*}
    \int_{\theta_{j-1}}^{\theta_j} f_{-2}(\theta)\frac{1}{\hat{\rho}(\theta)}  \ud \theta = \frac{f_{-2}^{(j)}}{h_j}\int_{\theta_{j-1}}^{\theta_j} \cos\theta \ud\theta.
\end{equation*}
The integral with $f_{-1}$ consists of terms
\begin{equation} \label{eq:I2j1}
    \ln h_j \int_{\theta_{j-1}}^{\theta_j} \cos^p\theta\sin^q\theta \ud\theta,  \qquad p+q=3 \quad \text{or} \quad p+q=1,
\end{equation}
and
\begin{equation}
    \bar{I}_i=\int_{\theta_{j-1}}^{\theta_j} \cos^p\theta\sin^q\theta\ln\cos \theta \ud t, \qquad p+q=3 \quad \text{or} \quad p+q=1.
\end{equation}
The explicit expressions of integral \eqref{eq:I2j1} can be easily obtained and are omitted. Here the expression of $\bar{I}_i$ is derived. By changing the integral variable
\begin{equation}
\begin{aligned}
    t&=\sin \theta\\
    \cos \theta&=\sqrt{1-t^2}\\
    \ud t&=\sqrt{1-t^2}\ud \theta,
\end{aligned}
\end{equation}
one obtains
\begin{equation}
\begin{aligned}
    \bar{I}_1&=\int_{\theta_{j-1}}^{\theta_j} \cos^3\theta\ln\cos\theta \ud \theta
        =\int_{t_{j-1}}^{t_j} t\sqrt{1-t^2}\ln\sqrt{1-t^2} \ud t = \tilde{I}_1-\tilde{I}_3,\\
    \bar{I}_2&=\int_{\theta_{j-1}}^{\theta_j} \cos^2\theta\sin\theta\ln\cos\theta \ud \theta
        =\int_{t_{j-1}}^{t_j} \frac{t-t^3}{\sqrt{1-t^2}}\ln\sqrt{1-t^2} \ud t = \tilde{I}_2-\tilde{I}_4,\\
    \bar{I}_3&=\int_{\theta_{j-1}}^{\theta_j} \cos\theta\sin^2\theta\ln\cos\theta \ud \theta
        =\int_{t_{j-1}}^{t_j} t^2\ln\sqrt{1-t^2} \ud t = \tilde{I}_3,\\
    \bar{I}_4&=\int_{\theta_{j-1}}^{\theta_j} \sin^3\theta\ln\cos\theta \ud \theta
        =\int_{t_{j-1}}^{t_j}\frac{t^3}{\sqrt{1-t^2}}\ln\sqrt{1-t^2} \ud t = \tilde{I}_4,\\
    \bar{I}_5&=\int_{\theta_{j-1}}^{\theta_j} \cos \theta\ln\cos \theta \ud \theta
        =\int_{t_{j-1}}^{t_j}\ln\sqrt{1-t^2} \ud t = \tilde{I}_1,\\
    \bar{I}_6&=\int_{\theta_{j-1}}^{\theta_j} \sin \theta\ln\cos \theta \ud \theta
        =\int_{t_{j-1}}^{t_j} \frac{t}{\sqrt{1-t^2}}\ln\sqrt{1-t^2} \ud t = \tilde{I}_2,
\end{aligned}
\end{equation}
where,
\begin{equation}
\begin{aligned}
    \tilde{I}_1&=\int_{t_{j-1}}^{t_j} \ln\sqrt{1-t^2}\ud t
        =\left. t\left( \ln\sqrt{1-t^2}-1\right) + \ln\frac{t+1}{\sqrt{1-t^2}} \right|_{t_{j-1}}^{t_j},\\
    \tilde{I}_2&=\int_{t_{j-1}}^{t_j} \frac{t\ln\sqrt{1-t^2}}{\sqrt{1-t^2}} \ud t
        =\left. \sqrt{1-t^2}\left(1-\ln\sqrt{1-t^2}\right) \right|_{t_{j-1}}^{t_j},\\
    \tilde{I}_3&=\int_{t_{j-1}}^{t_j} t^2\ln\sqrt{1-t^2} \ud t
        =\left. \frac{1}{9}t^3\left(3\ln\sqrt{1-t^2}-1\right)-\frac{1}{3}t+\frac{1}{6}\ln\frac{1+t}{1-t} \right|_{t_{j-1}}^{t_j},\\
    \tilde{I}_4&=\int_{t_{j-1}}^{t_j} \frac{t^3\ln\sqrt{1-t^2}}{\sqrt{1-t^2}} \ud t
        =\left. \sqrt{1-t^2}\left[\frac{1}{9}(8+t^2)-\frac{1}{3}(2+t^2)\ln\sqrt{1-t^2}\right] \right|_{t_{j-1}}^{t_j}.
\end{aligned}
\end{equation}

\section{Numerical examples} \label{examples}
\begin{figure}[htb]
    \centering
    \includegraphics[width = 0.3\textwidth]{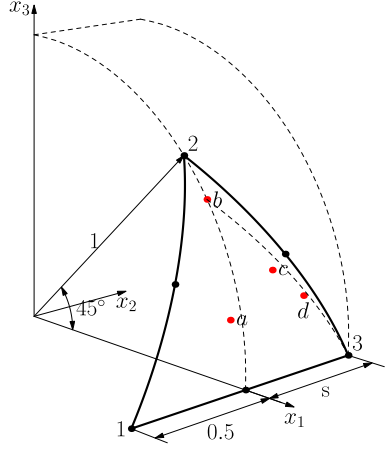}
    \caption{Curved element and the field points in example 1} \label{fig:ex1}
\end{figure}
Two numerical examples are provided to demonstrate the efficiency and robustness of the method in this paper.
The first one is used to illustrate the superior of the method in this paper to the method in \ref{s:polar_coord}, i.e., the classic polar coordinates transformation and Guiggiani's method \cite{Guiggiani}, using a sample curved element.
The second one is used to verify the accuracy and convergence of the method when employed in solving Burton-Miller equation. The method is implemented in C language. The numerical integration code are freely available.

The focus of this paper is to improve the performance of the quadrature in angular direction, while in radial direction,
as explained in section \ref{properties}, the integrand behaves well so that only a small number of quadrature points are sufficient.
In section \ref{example1} $6$ points are used in every test in radial direction.
However, when implemented in BEM programs, fewer points are needed because as the subdivision of the model
the diameter of each element will decrease so that fewer terms are needed in expansion \eqref{eq:expansion} to approximate the integrand.
As a result, the rate of convergence of BEM will not be affected. For the $2$th order Nystr\"om BEM in section \ref{ne:sound}, we use $3$ points in polar direction.

\subsection{Accuracy of the method on a curved element} \label{example1}
The accuracy, effectiveness and robustness of the method in this paper are investigated. The integrals is performed over a curved element extracted from a cylinder surface. The size of the cylinder and the element are given in Fig.\ref{fig:ex1}. Distance $s$ in Fig. \ref{fig:ex1} is designed to change the aspect ratio of the element. $s=0.5$ represents a quite regular element; the quality of the element becomes bad as the increase of $s$.
Four positions for the field point $\X$ are considered, which correspond to intrinsic coordinates (a) $\bm{\xi}=(0.3,0.3)$, (b) $\bm{\xi}=(0.1,0.8)$, (c) $\bm{\xi}=(0.45,0.45)$, (d) $\bm{\xi}=(0.64,0.31)$, respectively.
One should note that points (b) and (c) match approximately with the nodes of the $2$th order Nystr\"om discretization, while (d) match with that of the $3$th order. In addition, for the rather centered point (a) the original Guiggiani's method in section \ref{s:polar_coord} are believed to work well.

Since no analytical solutions are available, the (relative) error, given by
\begin{equation*}
    \text{relative error}=\frac{\norm{I_{\text{calc}}-I_{\text{ref}}}}{\norm{I_{\text{ref}}}}
\end{equation*}
is evaluated by comparison with the results $I_{\text{ref}}$ of the Guiggiani's methods using Gauss quadratures with $256$ quadrature points in angular direction and $32$ points in radial direction. $I_{\text{calc}}$ denotes the result calculated by the various methods.
The single layer operator $\mathcal{S}$ and the hypersingular operator $\mathcal{H}$ are tested. The parameter $m$ in sigmoidal transformation takes the value of $3$ in weakly singular case and $2$ in hypersingular case.
The basis function $\phi$ is selected as the second order monomial $\xi_2^2$.

\begin{table}
    \centering
    \caption{methods to compare} \label{tab:methods}
    \begin{tabular}[c]{ll}
        \hline
        methods         & description\\
        \hline
        Guiggiani       & Guiggiani's method in section \ref{s:polar_coord}\\
        Gui+sig             & Guiggiani's method with a sigmoidal transformation applied in angular direciton\\
        Present            & Method in this paper, i.e. Guiggiani's method with the two transformations in section \ref{integral}\\
        Present+a        & Similar to ``Present'', except that the line integral is calculated analytically\\
        \hline
    \end{tabular}
\end{table}

A description of the methods tested are listed in table \ref{tab:methods}. The method in this paper are indicated by ``Present'' in which the line integral $I_2$ in \eqref{eq:Guiggiani} is computed by Gauss quadratures; while in ``Present+a'' the line integral is computed by using the closed formulations in section \ref{analytic}. Note that ``Present+a'' only validates for hypersingular integrals.

\subsubsection{Results for regular element}

First the performance of the present method for regular element are verified; for this purpose, we let $s=0.5$ and the wave number $k=0$ (Laplace kernels).

\begin{figure}
    \centering
    \subfigure[field point (a)] {
        \includegraphics[width = 0.45\textwidth]{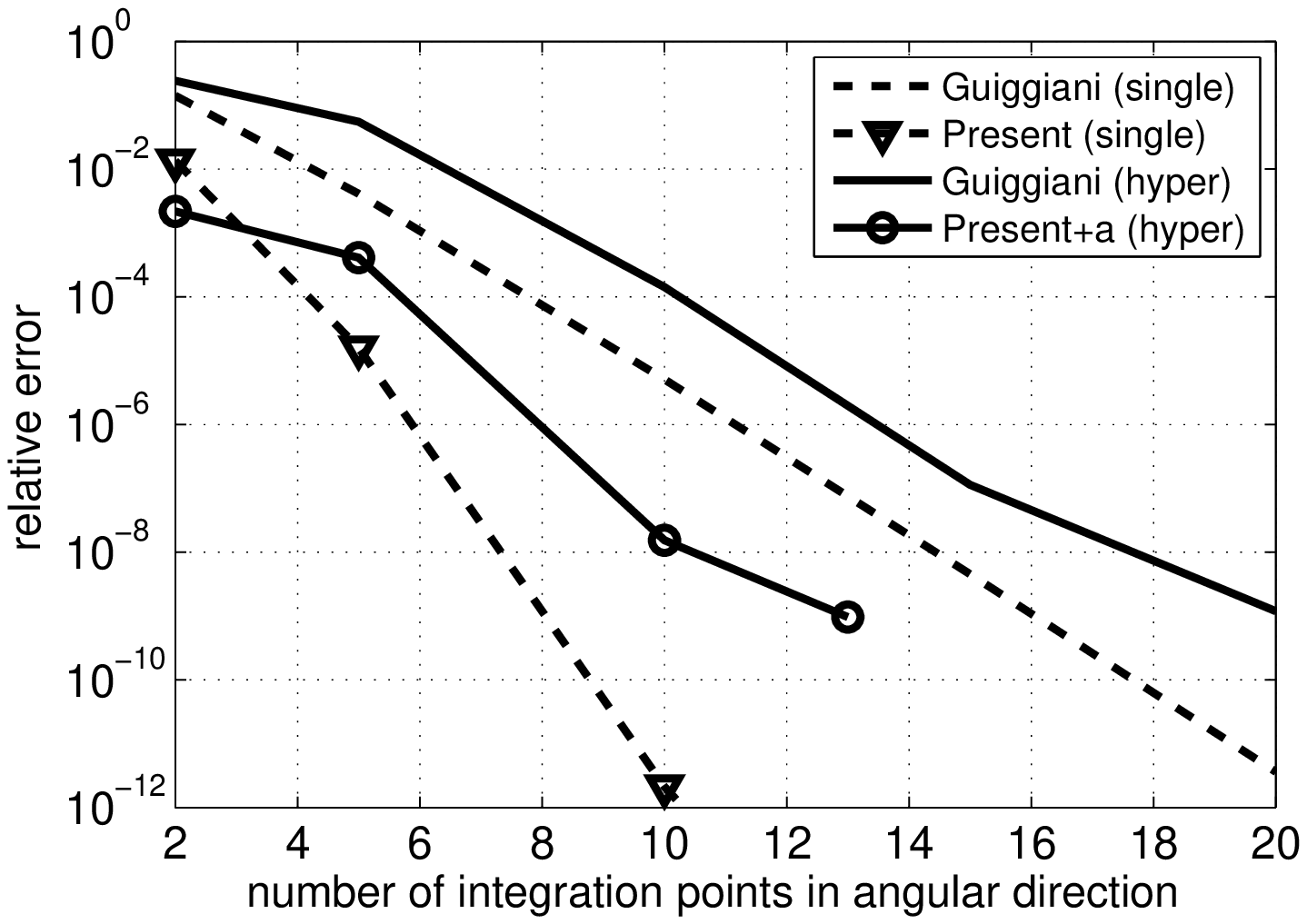}
    }
    \subfigure[field point (b)] {
        \includegraphics[width = 0.45\textwidth]{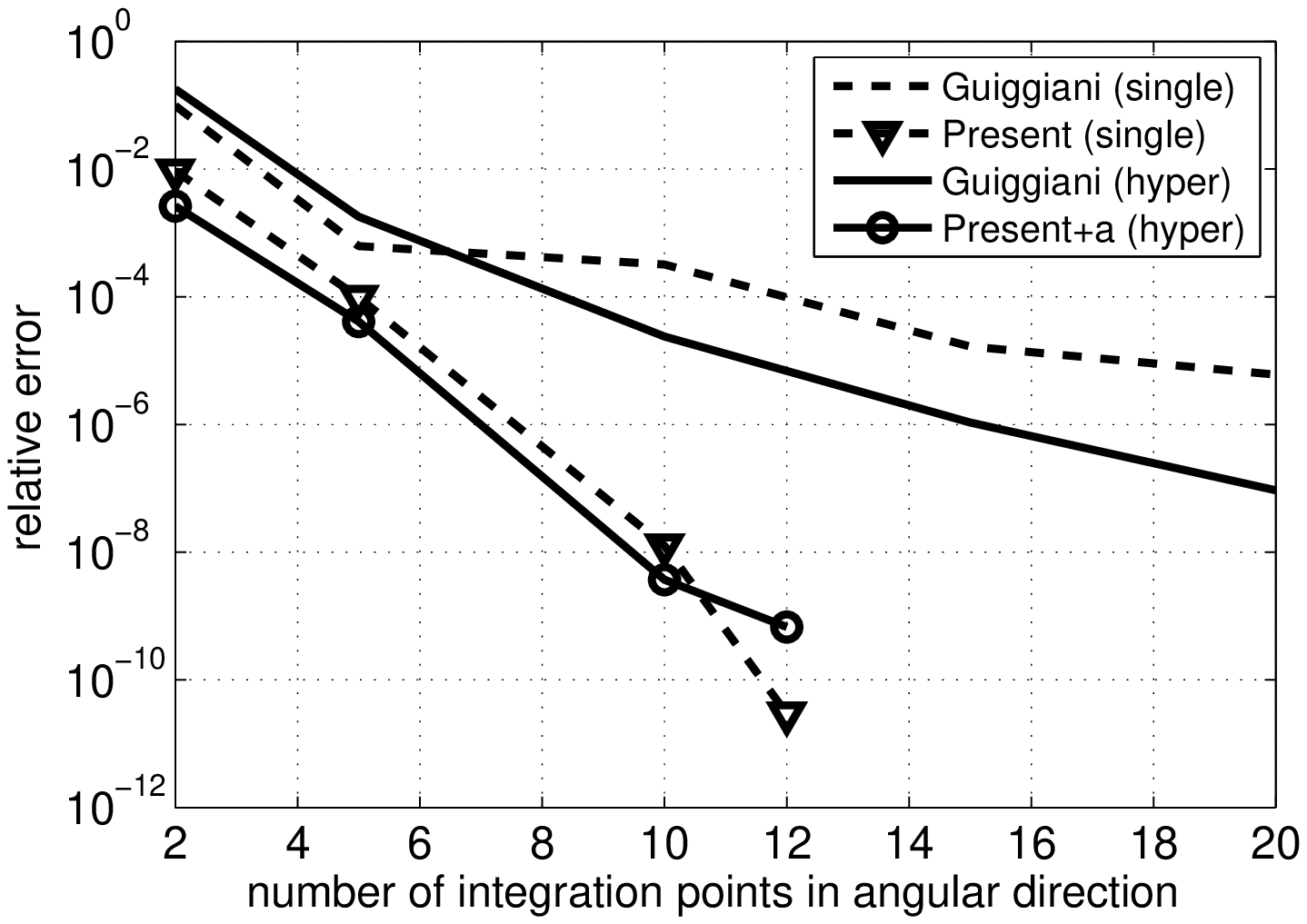}
    }\\
    \subfigure[field point (c)] {
        \includegraphics[width = 0.45\textwidth]{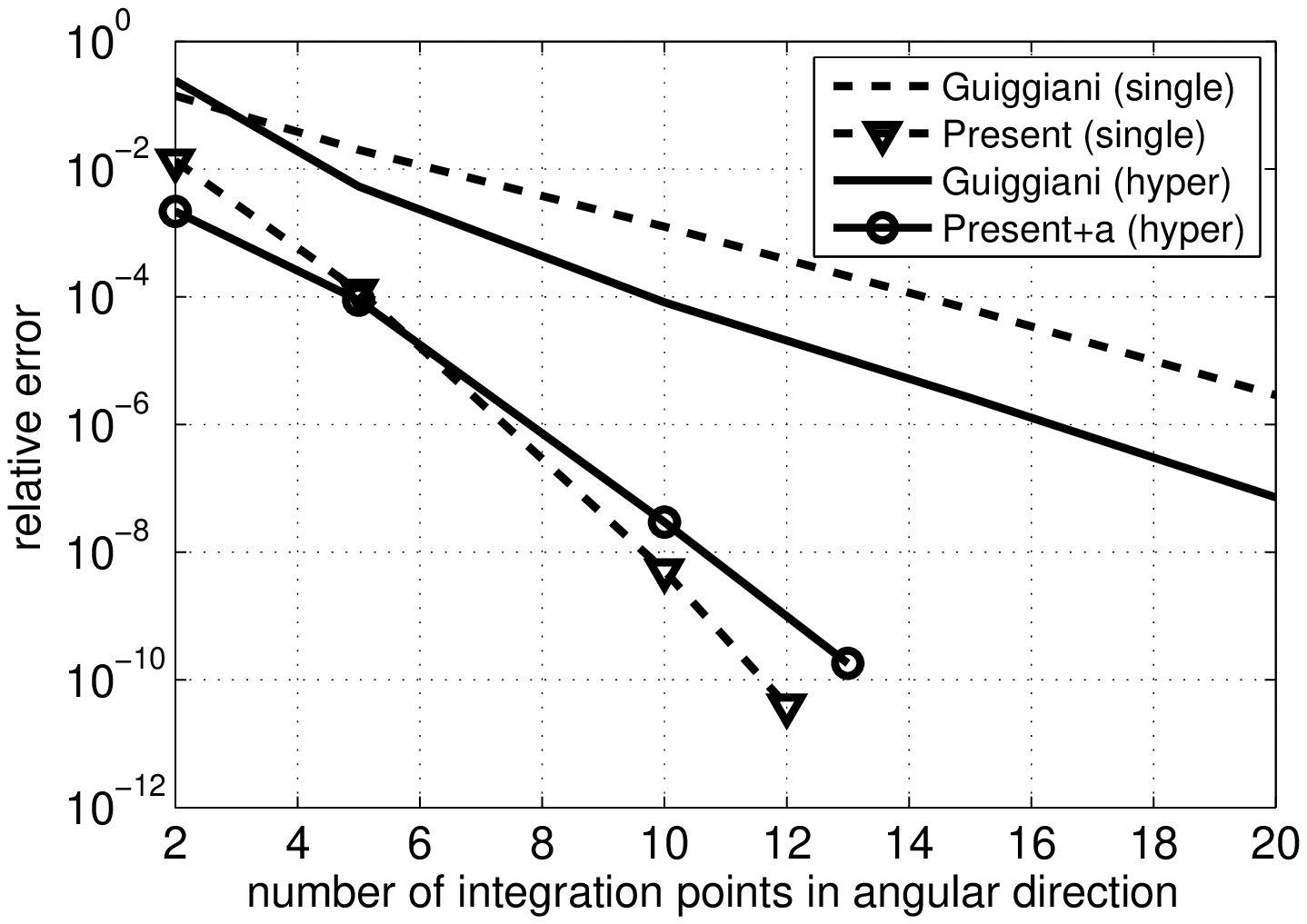}
    }
    \subfigure[field point (d)] {
        \includegraphics[width = 0.45\textwidth]{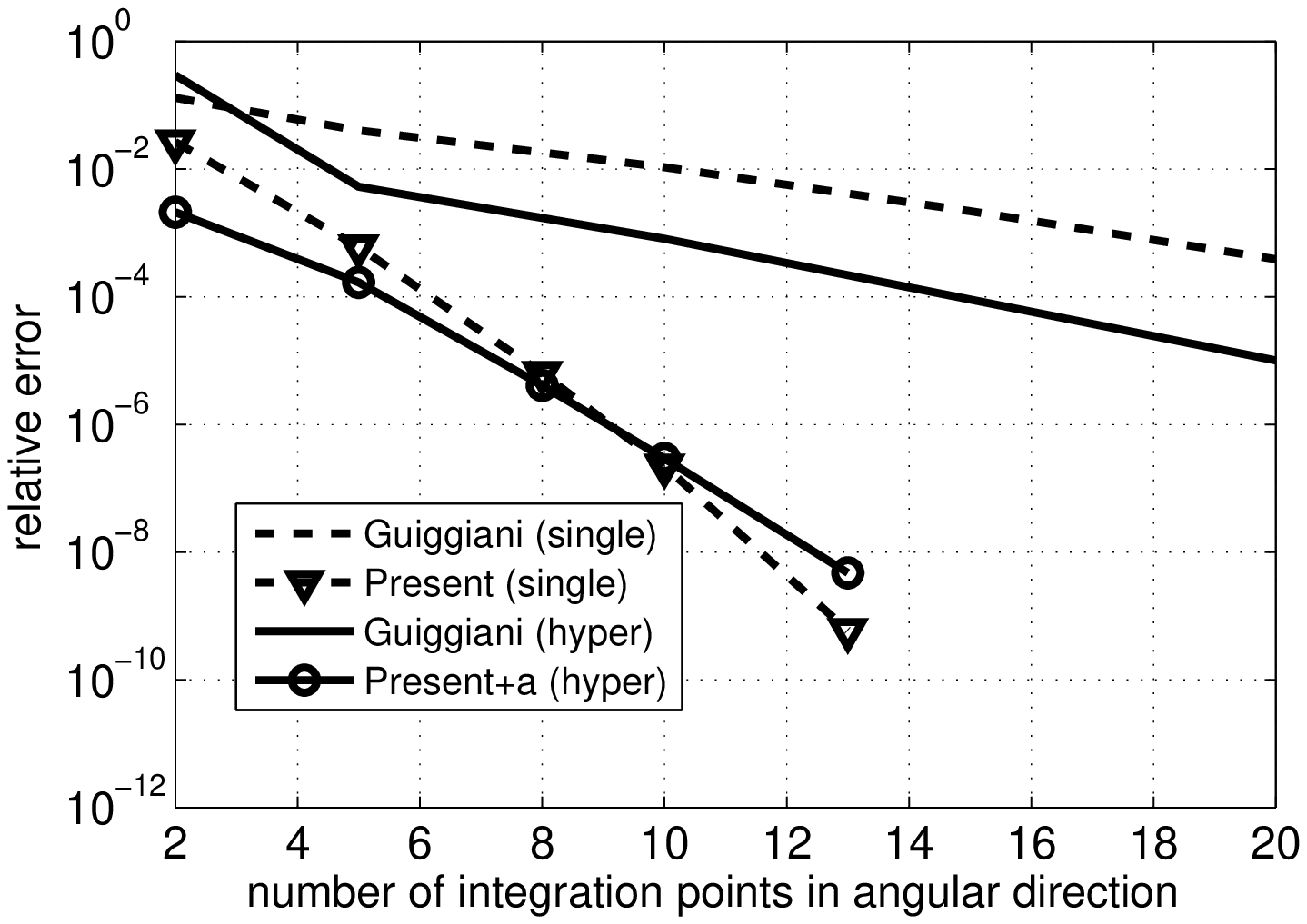}
    }
    \caption{Convergence comparison for the $k=0$ case.} \label{fig:error}
\end{figure}

Convergence behavior of the original methods and the present method for the four field points are illustrated in Fig.\ref{fig:error}.
The results of the present method are marked by triangles or circles. The results for single layer kernel are plotted with dashed line,
while real lines are for hypersingular integrals. A comparison between case (a) and the other three cases shows that the original Guiggiani's method tends to converge slowly when the field point is closed to the element boundary; while the present method can achieve almost the same fast convergence. Even in regular case (a) in which the original method performs better than other three cases, the benefit of the present method is also obvious.

The effect of each transformation scheme in section \ref{integral} are studied and exhibited in table \ref{tab:convergence}. Both field points (b) and (d) are tested. For the weak singular (single layer) integral, the accuracy can be improved by only using the sigmoidal transformation;
whereas the present method, say the combination of the two transformations in section \ref{integral}, can further promote the accuracy significantly. One notice that for the hypersingular case the sole use of the sigmoidal transformation can not improve the accuracy. However, this problem can be completely overcome by using the parametric coordinates transformation.
The analytical formulas for the line integrals can be better than the numerical quadratures.

Now consider the Helmholtz kernels. The results are similar to that of the Laplace kernels, so only the results for field point (b) are given in table \ref{tab:dynamic}. The wavenumber $k$ is selected as $2.0$ so that the wavelength is about $3$ times size of the triangular element. In practical BEM implementation such a mesh is rather coarse. It is obvious that the present method can greatly improve the accuracy.

\begin{table}
    \centering
    \caption{Relative errors for different methods. ``n'' denotes the number of Gauss quadrature points in angular direction.} \label{tab:convergence}
    \begin{tabular}[c]{c|ccc|cccc}
        \hline
                                & \multicolumn{3}{c}{single}         & \multicolumn{4}{c}{hyper}\\
        \hline
            n                   & Guiggiani  &Gui+sig        &Present      & Guiggiani    &Gui+sig       &Present    &Present+a\\
        \hline
                                                            &\multicolumn{7}{c}{field point (b)}\\
        \hline
            5                   & 6.24e-04  &1.22e-04   &9.83e-05    & 1.81e-03  &5.51e-03  &1.28e-03   &4.1e-05\\
            8                   & 1.02e-03  &2.83e-05   &6.44e-07    & 1.70e-04  &2.62e-04  &5.41e-06   &1.49e-07\\
            10                  & 3.20e-04  &3.08e-07   &1.26e-08    & 2.38e-05  &2.72e-05  &5.64e-08   &3.71e-09\\
            12                  & 5.65e-05  &8.91e-08   &2.95e-11    & 3.01e-06  &1.12e-05  &4.20e-10  &6.70e-10\\
        \hline
                                                            &\multicolumn{7}{c}{field point (d)}\\
        \hline
            5                   & 4.05e-02   &1.18e-03    &6.06e-04     &5.33e-03    &6.28e-03   &1.11e-02   &1.68e-04\\
            8                   & 1.89e-02   &3.20e-05    &6.35e-06     &2.35e-03    &1.48e-04   &7.32e-05   &4.08e-06\\
            10                  & 1.07e-02   &5.85e-07    &2.23e-07     &8.11e-04    &2.65e-04   &1.93e-06   &2.92e-07\\
            12                  & 5.83e-03   &1.50e-08    &5.16e-09     &3.45e-04    &3.15e-05   &5.34e-08   &1.57e-08\\
        \hline
    \end{tabular}
\end{table}

\begin{table}
    \centering
    \caption{Relative errors in dynamic case with $k=2.0$, the field point is selected as (b).}\label{tab:dynamic}
    \begin{tabular}[c]{c|cccc}
    \hline
                 & \multicolumn{2}{c}{single}         & \multicolumn{2}{c}{hyper}\\
    \hline
        n        & Guiggiani    & Present               & Guiggiani     & Present+a\\
    \hline
        5        & 2.52e-03     & 2.00e-04              & 1.92e-03      & 3.64e-04\\
        8        & 1.16e-03     & 2.44e-06              & 2.08e-04      & 3.22e-07\\
        10       & 4.09e-04     & 1.09e-07              & 2.04e-05      & 6.93e-09\\
        12       & 1.26e-04     & 4.66e-09              & 4.36e-06      & 4.71e-10\\
    \hline
    \end{tabular}
\end{table}
\subsubsection{Results for distorted element}

As have been mentioned in section \ref{change}, the present method should be robust for elements with high aspect ratio; here this is validated.
The integrals are evaluated over elements with different $s$ and $k=0$.
Point (a) is taken to be the field point. Both the original and present methods are used to compute the integrals with relative error $10^{-8}$. The needed number of Gauss quadrature points in angular direction are listed in table \ref{tab:number}.
It is seen that this number of the original method increases drastically as the increase of $s$, which clearly indicates that the performance of the original method depend heavily on the shape of the element. On the contrary, the present method behaves more robust with the change of $s$. For element with aspect ratio larger than $10$, high accurate results can still be achieved with a moderate increase of quadrature points. %Similar to the results in regular element case, the analytical evaluation of line integral

\begin{table}
    \centering
    \caption{Number of integration points needed in angular direction to make the relative error under $10^{-8}$} \label{tab:number}
    \begin{tabular}[c]{c|ccccc}
    \hline
        \multirow{2}{*}{s} & \multicolumn{2}{c}{single} & \multicolumn{3}{c}{hyper}\\
                            & Guiggiani  & Present          & Guiggiani & Present & Present+a\\
    \hline
        0.5                 & $15$       & $8$              & $18$     & $11$     & $11$\\
        1.5                 & $19$       & $9$              & $27$     & $12$     & $11$\\
        2.0                 & $22$       & $9$              & $31$     & $13$     & $12$\\
        4.0                 & $42$       &$11$              & $62$     & $14$     & $13$\\
        10.0                & $82$       & $13$             &$142$   & $15$     & $16$\\
    \hline
    \end{tabular}
\end{table}

\subsection{An exterior sound radiation problems} \label{ne:sound}
\begin{figure}
    \centering
    \includegraphics[width = 0.4\textwidth]{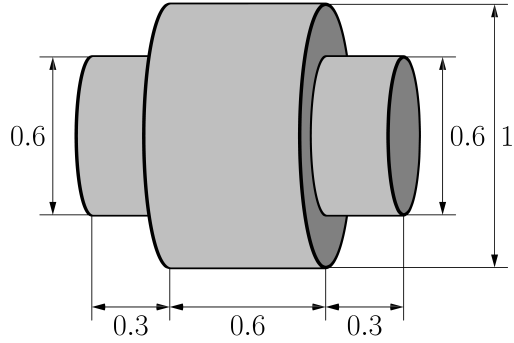}
    \caption{Geometry model}\label{fig:muffler}
\end{figure}

The overall performance of the present integration method is tested by incorporating it into a Nystr\"om BEM code for solving the Burton-Miller equation. The geometry of the problem consists of three sections of cylinders, as shown in Fig. \ref{fig:muffler}. The origin of coordinate system lies on the middle point of the symmetry axis of the cylinder. All the surfaces of the cylinder vibrate with given velocity $q(\X)=\partial u/ \partial \X$ (Neumann problem), where $u$ is chosen to be the fundamental solution $u(\X) = G(\X,\Y)$ in \eqref{eq:fundsol} with $\Y = (0,0,0)$.

The Nystr\"om BEM with $2$th order basis functions is used to discretize Burton-Miller equation. The wave number $k=2.5$. The surface is meshed by quadratic triangular element and refined four times; the finest mesh consists of $1638$ elements. %, thus there are around 3 elements in a wavelength.
In evaluating the singular integrals, both the original Guiggiani's method and the present method are tested. A common $m=2.5$ is used in the sigmoidal transformation for all the singular integrals, and $3$-point Gauss quadrature are used in radial direction.

\begin{figure}
    \centering
    \includegraphics[width = 0.55\textwidth]{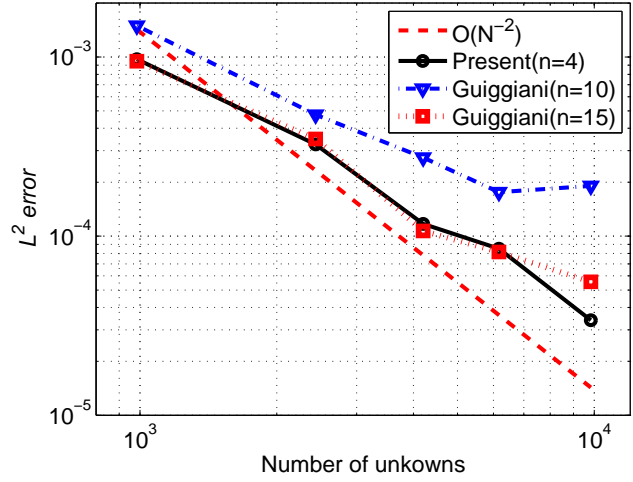}
    \caption{$L^2$-errors of $u$ in example \ref{ne:sound}.}\label{fig:error_BEM}
\end{figure}

Three different quadratures in the angular direction are used and compared, that is, the present method using $4$ points and the Guiggiani's method with $10$ and $15$ points. The $L^2$ relative errors of the boundary values of $u(\X)$ are demonstrated in Fig. \ref{fig:error_BEM}. It is shown that the accuracy of BEM is retarded when using the Guiggiani's method with 10 points. For the Guiggiani's method with 15 points, the convergence of the BEM tends to slow down with further refinement of the mesh. The present method with $4$ points, however, can achieve the theoretical converge rate.

\section{Conclusions} \label{conclusions}

Highly efficient methods with high accuracy and low computational cost are crucial for high-order boundary element analysis.
In \cite{Guiggiani}, Guiggiani proposed an unified framework to treat the singular integrals of various orders in BEM. It is based on the polar coordinate transformation which has been extensively used in dealing with BEM singular integrals. However, the performance of the polar coordinate transformation deteriorates when the field point is close to the element boundary or the aspect ratio of the element becomes large.

In this paper, first, a conformal transformation is introduced to circumvent the near singularity caused by large aspect ratio of element.
This transformation maps a curved physical element onto a planar triangle. Since it is conformal at the field point, the resultant integration domain (planar triangle) perseveres the shape of the curved element. Then, a sigmoidal transformation is applied to alleviate the near singularity due to the closeness of the field point to the element boundary. The rationale behind this is that the sigmoidal transformation can cluster the quadrature points to area of near singularity. The combination of the two transformations can effectively alleviate the two problems of the existing polar coordinate transformation method, and thus leads to considerable reduction of quadrature points in angular direction.

The efficiency and robustness of the present method are illustrated by various singular integrals on a curved quadratic element which is typical
in high order Nystr\"om BEM. It is shown that highly accurate results with relative error $10^{-8}$ can be achieved with $10$ quadrature points in angular direction. Moreover, the method is more stable with the change of element aspect ratio. For further verification, the method is applied to a $2$-order Nystr\"om BEM for solving acoustic Burton-Miller equation. Theoretical convergence rate of the BEM can be retained with much less quadrature points than the existing quadrature methods.

\section*{Acknowledgements}

This work was supported by National Science
Foundations of China under Grants 11074201 and 11102154 and Funds for Doctor Station from the Chinese Ministry of
Education under Grants 20106102120009 and 20116102110006.

\appendix
\section{Coefficients in Eq. \eqref{eq:f1}}\label{App:coefficient}

Consider a general integrand in BEM in polar coordinates
\begin{equation}\label{eq:integrand}
    F(\rho,\theta)=\frac{\rho\bar F(\rho,\theta)}{r^\beta},
\end{equation}
where, $\beta$ is the order of singularity, $\bar F(\rho,\theta)$ is a regular function.
There holds
\begin{equation} \label{eq:exp_r}
    \frac{1}{r^\beta}=\rho^{-\beta}\sum_{\nu=0}^{\infty}S_{\nu-\beta}(\theta)\rho^{\nu},
\end{equation}
\begin{equation}\label{eq:exp_F}
    \bar F(\rho,\theta)=\sum_{\nu=0}^{\infty}a_v(\theta)\rho^{\nu},
\end{equation}
where, $S_{\nu}(\theta)$ is given by (\cite{kernel}, Theorem 4)
\begin{equation}\label{eq:exp_S}
    S_{\nu-\beta}(\theta)=A^{-\beta-2\nu}(\theta)g_{3\nu}(\theta),
\end{equation}
$g_{3\nu}(\theta)$ are homogeneous trigonometric polynomials of order $3v$.

For hypersingular integrand, $\beta=3$, thus,
\begin{equation}\label{eq:f}
    \begin{aligned}
    f_{-2}(\theta)&=S_{-3}(\theta)a_{0}(\theta),\\
    f_{-1}(\theta)&=S_{-2}(\theta)a_{0}(\theta)+S_{-3}(\theta)a_{1}(\theta).
\end{aligned}
\end{equation}
See \cite{Guiggiani} for the expressions of $S_{-2}(\theta)$ and $S_{-3}(\theta)$.
%with \cite{Guiggiani}
%\begin{equation}
%\begin{aligned}
%    S_{-3}=&\frac{1}{A^3(\theta)},\\
%    S_{-2}=&-\frac{3 A_kB_k}{A^5(\theta)}
%\end{aligned}
%\end{equation}

For hypersingular kernel of Burton-Miller equation in this paper, the expressions $a_{i}$ can be obtained similarly with \cite{Guiggiani}. Let $J_k$ be the $k$th component of the vector
\begin{equation*}
    \pfrac{\Y}{\eta_1}\times\pfrac{\Y}{\eta_2}.
\end{equation*}
Then $J_k$ can be expanded as
\begin{equation*}
    J_k=J_{k0}+\rho\left[ \left.\pfrac{J_k}{\eta_1}\right|_{\bm{\eta}=\bm{\eta}^s}\cos\theta+
    \left.\pfrac{J_k}{\eta_2}\right|_{\bm{\eta}=\bm{\eta}^s}\sin\theta \right] =J_{k0}+\rho J_{k1}(\theta)+O(\rho^2).
\end{equation*}
The basis function $\phi$ can be expanded analogously as
\begin{equation*}
    \phi=\phi_0+\rho\left[ \left.\pfrac{\phi}{\eta_1}\right|_{\bm{\eta}=\bm{\eta}^s}\cos\theta+
    \left.\pfrac{\phi}{\eta_2}\right|_{\bm{\eta}=\bm{\eta}^s}\sin\theta \right]=\phi_0+\rho\phi_1 (\theta) +O(\rho^2).
\end{equation*}
Then, $a_0$ and $a_1(\theta)$ in Eq. \eqref{eq:f} can be expressed as (repeated indicies imply summation)
\begin{equation}\label{eq:a}
\begin{aligned}
    a_0 &= \frac{n_i(\X)}{4\pi}J_{i0}\phi_0,\\
    a_1(\theta)&=\frac{n_i(\X)}{4\pi}\left[ J_{i1}(\theta)\phi_0+J_{i0}\phi_1(\theta) \right].
\end{aligned}
\end{equation}

Substituting Eq. \eqref{eq:a} into Eq. \eqref{eq:f}, yields
\begin{equation}\label{eq:A-f1}
    \begin{split}
    f_{-1}(\theta)=& c_1\cos^3\theta+c_2\cos^2\theta\sin\theta+c_3\cos\theta\sin^2\theta+c_4\sin^3\theta\\
                    &+d_1\cos\theta+d_2\sin\theta,\\
   f_{-2}(\theta)=& \frac{n_i(\X)}{4\pi A^3}J_{i0}\phi_0.
    \end{split}
\end{equation}
with
\begin{equation}\label{eq:aibi}
\begin{aligned}
    c_1&=-\frac{3n_i(\X)J_{i0}\phi_0}{8\pi A^5}\pfrac{y_k}{\eta_1}\ppfrac{y_k}{\eta_1},\\
    c_2&=-\frac{3n_i(\X)J_{i0}\phi_0}{4\pi A^5}\left( \pfrac{y_k}{\eta_1}\ppfracd{y_k}{\eta_1}{\eta_2} + \frac{1}{2}\pfrac{y_k}{\eta_2}\ppfrac{y_k}{\eta_1} \right),\\
    c_3&=-\frac{3n_i(\X)J_{i0}\phi_0}{4\pi A^5}\left( \pfrac{y_k}{\eta_2}\ppfracd{y_k}{\eta_1}{\eta_2} + \frac{1}{2}\pfrac{y_k}{\eta_1}\ppfrac{y_k}{\eta_2} \right),\\
    c_4&=-\frac{3n_i(\X)J_{i0}\phi_0}{8\pi A^5}\pfrac{y_k}{\eta_2}\ppfrac{y_k}{\eta_2},\\
    d_1&=\frac{n_i(\X)}{4\pi A^3}\left( \pfrac{J_i}{\eta_1}\phi_0 + J_{i0}\pfrac{\phi}{\eta_1} \right),\\
    d_2&=\frac{n_i(\X)}{4\pi A^3}\left( \pfrac{J_i}{\eta_2}\phi_0 + J_{i0}\pfrac{\phi}{\eta_2} \right).
\end{aligned}
\end{equation}
Note that all the derivatives are evaluated at the field point $\X$.
%\subsection{Strong singular kernel in elastic CBIE}
%\begin{equation}
%    a_0
%\end{equation}
%\subsection{Hypersingular kernel in elastic HBIE}

\bibliographystyle{model1-num-names}

\end{document}